\documentclass{ws-m3as}

\usepackage{graphicx}       
\usepackage{xcolor}         
\usepackage{overpic}        
\usepackage{enumitem}

\usepackage{bm}             
\usepackage[ruled,linesnumbered]{algorithm2e} 

\usepackage{tikz}
\usetikzlibrary{shapes,arrows,decorations.markings,decorations.pathmorphing}

\usepackage[mathlines]{lineno} 
\usepackage{soul}              

\usepackage[pdftex]{hyperref}

\tikzstyle{block_long} = [rectangle, draw, fill=blue!20,
    text width=10.0em, text centered, rounded corners, minimum height=3em]
\tikzstyle{block_medium} = [rectangle, draw, fill=blue!20,
    text width=6.0em, text centered, rounded corners, minimum height=3em]
\tikzstyle{block} = [rectangle, draw, fill=blue!20,
    text width=3.0em, text centered, rounded corners, minimum height=3em]
\tikzstyle{line} = [thick, draw, dashed,  -stealth']

\hypersetup{
  colorlinks=true,     
  linkcolor=blue,      
  citecolor=blue,      
  urlcolor=magenta,    
}

\makeatletter
\renewcommand\LineNumber{\hb@xt@\linenumberwidth{\hss\normalfont\footnotesize\arabic{linenumber}}}
\makeatother

\def\bv{{\mathbf v}}

\everymath{\displaystyle}

\begin{document}

\markboth{Fabregas, Liao \& Outada}{Mathematical Theory of Behavioural Swarms}

%
\catchline{}{}{}{}{}
%

\title{The Mathematical Theory of Behavioural Swarms: Towards Modelling the Collective Dynamics of Living Systems}

\author{Rene Fabregas}

\address{Department of Applied Mathematics and Modeling Nature (MNat) Research Unit,\\
Faculty of Sciences, University of Granada, Granada, Spain\\
rfabregas@ugr.es}

\author{Jie Liao}

\address{School of Mathematics,\\
Shanghai University of Finance and Economics,\\
Shanghai 200433, P. R. China\\
liaojie@mail.shufe.edu.cn}

\author{Nisrine Outada}

\address{Laboratory of Mathematics, Modeling and Automatic Systems,\\
University Cadi Ayyad (UCA),\\
 Faculty of Sciences Semlalia, Marrakesh, Morocco\\
UMMISCO, IRD-SU, France\\
nisrine.outada@uca.ac.ma}

\maketitle

\begin{history}
\received{(Day Month Year)}
\revised{(Day Month Year)}
\comby{(xxxxxxxxxx)}
\end{history}

\begin{abstract}
Classical swarm models, exemplified by the Cucker--Smale framework, provide foundational insights into collective alignment but exhibit fundamental limitations in capturing the adaptive, heterogeneous  behaviours intrinsic to living systems. This paper  formalises the mathematical theory of \textit{Behavioural Swarms}, a comprehensive framework where each particle's state incorporates a dynamic internal variable, the \textit{activity} that co-evolves with position and velocity through nonlocal interactions. We demonstrate how this approach transcends prior models by integrating adaptive decision-making mechanisms and heterogeneous  behavioural states into rigorous differential systems. Through applications in  behavioural economics and crowd dynamics, we establish the theory's capacity to predict emergent macroscopic patterns from individual  behavioural states. Our critical analysis positions this framework against kinetic theories of active particles and agent-based approaches, revealing distinct advantages for  modelling systems where individual agency drives collective outcomes.
\end{abstract}



\section{Conceptual foundations and exposition}\label{sec:1}
One of the most profound challenges in modern science lies in developing a mathematical language capable of describing the dynamics of living systems~\cite{[Schrodinger],[vonNeumann66]}. Unlike inert matter, living entities exhibit adaptive strategies, heterogeneous behaviours, and a capacity for self-organisation that defy classical mechanical descriptions. This quest for a ``calculus of life'' has motivated intense research, from the foundational principles of causality sought by Schr\"odinger~\cite{[Schrodinger]} to the statistical physics of complex systems pioneered by Prigogine~\cite{[Ilia]} and Parisi~\cite{[Parisi23]}. Nobel laureate Lee Hartwell voiced this challenge explicitly, urging mathematicians to move beyond frameworks applicable to inert matter and develop new theories specifically for living systems~\cite{[Hartwell]}. This paper addresses this call by formalising and critically reviewing the mathematical theory of \textit{Behavioural Swarms Theory (BST)}, a framework designed to bridge the gap between mechanical motion and the internal, non-physical states that govern the behaviour of living particles.

The study of collective motion has been dominated by seminal models that laid the groundwork for the field. The Cucker-Smale (CS) model, for instance, provides a foundational description of flocking based on velocity alignment, where particles adjust their speed and direction according to a weighted average of their neighbours' velocities~\cite{[CS07],[CS2007]}. Parallel to this, the Vicsek model offered a simpler yet profound paradigm based on directional alignment among particles moving at a constant speed~\cite{[Vicsek95]}. Extensions incorporating network topologies and coordination algorithms~\cite{[Jadbabaie03]} have further enriched this framework by addressing consensus dynamics in distributed systems. These models, and their numerous extensions, have been instrumental in explaining emergent synchronisation and pattern formation. However, they fundamentally treat particles as automata governed by predetermined rules based on metric or topological distance~\cite{[Ballerini]}. This abstraction, while potent, overlooks a defining characteristic of living systems: the ability of individuals to modulate their behaviour based on internal states such as fear, stress, confidence, or motivation.

To overcome this limitation, the BST proposes a necessary paradigm shift. It enriches the state of each interacting particle—henceforth called an \textit{active particle (a-particle)}—with a dynamic internal variable, termed \textit{activity}. This variable is not a fixed parameter but a co-evolving component of the system's state, heterogeneously distributed and modified by interactions. The core hypothesis of BST is that a reciprocal feedback loop exists: activity influences the mechanical dynamics (e.g., alignment, speed), while the collective motion, in turn, shapes the evolution of activity. This concept builds upon a rich history of modelling efforts, from the ``social forces'' in pedestrian dynamics~\cite{[Helbing95]} to the kinetic theory of active particles (KTAP), which has been successfully applied to model complex biological and social systems~\cite{[BBGO17],[BBD21]}. Indeed, the BST framework can be viewed as a specific, mathematically rigorous formulation within the broader landscape of ``active matter'' physics, a field dedicated to systems whose constituent particles consume energy to produce directed motion~\cite{[Marchetti13]}. Recent advances in active matter theory~\cite{[Toner95],[Couzin06],[Cavagna10]} have revealed universal scaling laws and critical phenomena in biological swarms.

This paper provides a review and critical analysis of the BST, framing it as a natural and substantial development of classical swarm theory. While other mathematical tools have been developed to tackle multi-agent systems---notably the theory of mean-field games~\cite{[Bartucci25],[LL07]} and recent advances in optimal-control theory for multi-agent systems~\cite{[Caponigro17]}, as well as kinetic descriptions based on the Fokker–Planck–Boltzmann equation~\cite{[PT13]}---our focus remains on the differential-equation-based framework of BST. This approach offers a means of analysing system stability, bifurcations, and the explicit connection between microscopic interaction rules and macroscopic emergent behaviours. We aim to demonstrate how this theory can describe the dynamics of systems composed of interacting living entities, providing a unified structure that can be specialised to a vast range of applications, from bacterial colonies~\cite{[Ben-Jacob00]} and bird flocks~\cite{[Vicsek95]} to social sciences and behavioural economics~\cite{[BBST24],[KST20]}. Building on this critical review and comparing it with similar methods, we examine future prospects, including applications to swarm intelligence~\cite{[BDL24],[KKHP]} and collaborations with scientific machine-learning frameworks~\cite{[Karniadakis21]}. Specifically, we will outline a future direction for BST by utilising modern techniques for system identification and data-driven simulation, such as Sparse Identification of Nonlinear Dynamics (SINDy)~\cite{[Brunton2016pnas]} and Physics-Informed Neural Networks (PINNs)~\cite{[Raissi2019jcp]}.

Section 2 provides an overview of CS models, beginning with a description of the seminal paper~\cite{[CS07]}, see also~\cite{[CS2007]}.  It then reviews several developments relevant to living systems, namely hierarchical leadership, herding behaviour in financial markets, models with internal thermodynamic variables, and stochastic variants with environmental noise. These topics motivate the need for a new swarm theory capable of capturing the complex features of living systems

Section 3 introduces the mathematical theory of behavioural swarms consisting of interacting active particles, which represent individual living entities. This theory is somewhat related to the mathematical kinetic theory of active particles; see~\cite{[BBD21]}. This section reports on the seminal paper~\cite{[BHO20]} and on some subsequent developments of the theory that have been motivated by applications; see~\cite{[BHLY24],[BHOY22]}.

Section 4 begins with a review of two applications of behavioural-swarm theory, emphasising the influence of heterogeneous individual behaviours on swarm dynamics. The first case study focuses on the role of activity (stress) in shaping swarm behaviour~\cite{[BHO20],[BHOY22]}, and the second focuses on the collective dynamics of price series in a market~\cite{[KST20]}. Next, numerical studies are proposed to demonstrate the influence of {\it interaction domain} on panic propagation with different visibility angles, and {\it multi-consensus} dynamics in behavioural swarms.

Section 5 looks ahead to future research perspectives. First, we critically analyse the advantages and disadvantages of the BST in relation to the kinetic theory of active particles, based on the content of the previous sections. This preliminary study enables us to consider applications requiring further development of a mathematical theory, to which we draw the attention of interested readers. Lastly, we explore developments related to swarm intelligence and scientific machine learning.

\section{Cucker--Smale flocking model}\label{sec:2}

The BST is a development of the mathematical theory of swarms introduced by Cucker and Smale in their seminal paper~\cite{[CS07]}, in which the dynamical Cucker--Smale (CS) model was proposed to study asymptotic swarm dynamics in a population of interacting particles, focusing on the influence of initial configurations and inter-agent communication.

The seminal work of Cucker and Smale has generated significant research interest, resulting in numerous extensions to the original concept. These developments have improved our understanding of flocking phenomena and making the model useful in a wide range of real-world situations. In particular, the CS model has become a valuable tool for studying collective behaviour in various contexts.

Although this review focuses on the BST, it is useful to provide a brief overview of the CS model in order to clarify the underlying conceptual framework. 

This section focuses on the original model and some specific developments that may contribute to the advancement of BST. These issues are presented in the following subsections, which cover the following topics: a brief presentation of the original CS work; four models involving different types of internal organisation—namely, models under hierarchical leadership~\cite{[LX10],[S2008]}, models describing the herding behaviour of particles in financial markets~\cite{[BCKY19],[BCLY19]}, the thermodynamic Cucker--Smale model~\cite{[HKR18],[HR17]}, and stochastic models~\cite{[HLL09]}. Interesting contributions to the search for new interaction rules are given in~\cite{[MT11B]}, while~\cite{[CK02]} offers insights into how memory effects influence pattern formation in collective behaviour.

Readers interested in a deeper study of the mathematical theory of swarms are referred to the review paper~\cite{[ALBI19]}, see also~\cite{[MT11]}. These two papers present various aspects of the theory, including its interaction with the mathematical theory of games~\cite{[Nowak],[NM01]}.

We also refer readers to the comprehensive review~\cite{[CHL17]}, which provides a detailed overview of Cucker--Smale-type flocking models. These include microscopic, mesoscopic and macroscopic descriptions, as well as rigorous flocking theorems under various symmetric and non-symmetric interaction topologies. The review also discusses extensions involving bonding forces, discrete-time models featuring leadership structures (hierarchical, alternating and rooted) and kinetic–fluid coupling models.

This section begins with a technical description of the Cucker--Smale model. It then provides a brief bibliography of papers considering interesting aspects of living systems and motivates the need to move to the theory of behavioural swarms. These topics are covered in the subsequent subsections.

\subsection{The Cucker--Smale model}\label{sec:2.1}

Consider a flock consisting of $N$ particles, indexed by $i = 1, 2, \ldots, N$, moving in $\mathbb{R}^d$, where $d = 2$ or $3$. The state of each particle is described by its position $\bm{x}_i(t) \in \mathbb{R}^d$ and velocity $\bm{v}_i(t) \in \mathbb{R}^d$.

This section reports on the Cucker--Smale model, in which each particle adjusts its velocity based on the velocity differences between itself and the other particles using a weighted average. This rule reflects the natural tendency of particles to align their movements. The model is expressed mathematically as follows:
\begin{equation}\label{eq:cs-velocity}
\frac{d \bm{v}_i}{dt} = \frac{1}{N}\sum_{j=1}^N a_{ij}\{\bm{x}_i, \bm{x}_j\} (\bm{v}_j - \bm{v}_i),
\end{equation}
where $a_{ij}$ represents the communication coefficients that quantify the degree of interaction between particles, while the brackets are used to distinguish functional dependence. These coefficients typically depend on factors such as the distance between particles $i$ and $j$ and are given by
\begin{equation}\label{eq:connectivity-matrix}
a_{ij} = \psi(\|\bm{x}_i - \bm{x}_j\|^2),
\end{equation}
where the standard Euclidean norm in $\mathbb{R}^d$ is denoted by $\|\cdot\|$, and the function $\psi \colon \mathbb{R}_+ \longrightarrow \mathbb{R}_+$ is a non-negative communication weight function. The choice of the weight function is a key aspect that makes the CS model particularly attractive. In the original model~\cite{[CS07]}, the weight function is specified as follows:
\begin{equation}\label{eq:cucker-smale-communication}
\psi(r) = \frac{K}{(1+r)^\beta}, \quad  r \in \mathbb{R}_+,
\end{equation}
where $K$ and $\beta$ are positive parameters and $\psi(r)$ implies that the intensity of interactions decays with distance $r$, capturing the tendency of particles to exert a stronger influence on neighbours in closer proximity.

The velocity-alignment equation \eqref{eq:cs-velocity} is combined with another equation describing how the positions of the particles change over time. Together, these two equations form the complete CS model describing particle motion in a pseudo-Newtonian framework:
\begin{equation}\label{eq:cucker-smale-model}
\begin{cases}
\frac{d \bm{x}_i}{dt} = \bm{v}_i,\\[3mm]
\frac{d \bm{v}_i}{dt} = \frac{1}{N} \sum_{j=1}^N a_{ij}\{\bm{x}_i, \bm{x}_j\} (\bm{v}_j - \bm{v}_i).
\end{cases}
\end{equation}

Here, we summarise the key results, which demonstrate that the system's flocking behaviour depends on the parameter $\beta$:
\begin{itemize}[itemsep=2mm]
\item[(1)] When $\beta < 1/2$: The system always achieves global flocking, meaning that all particles will eventually move together at the same constant velocity, regardless of their initial positions and velocities.
\item[(2)] When $\beta \geq 1/2$: Flocking is still possible, but only if the initial conditions meet specific requirements.
\end{itemize}

These results highlight the role of $\beta$ in determining how the system behaves. When $\beta$ is small ($< 1/2$), long-range interactions dominate, leading to strong synchronisation among particles. In contrast, when $\beta$ is larger ($\geq 1/2$), the initial conditions become more critical, making flocking less predictable and more sensitive to variations.

While the rigorous mathematical analysis of the canonical CS model is laid out in their seminal paper \cite{[CS07]}, it is in the subsequent extensions—which we explore next—that the framework begins to confront the complexities of real-world living systems.

\subsection{Some developments of the CS model}\label{sec:2.2}
The original CS model was briefly described in the previous section. In view of the study proposed in the next sections, it is useful to provide a brief description of some technical modifications to the model, focusing on those that relate to the main content of this paper, which is dedicated to searching for differential tools to describe the dynamics of living systems. In particular, the following classes of models have been selected because they incorporate interaction structures that reflect features typical of living systems and aspects of human behaviour. However, it should be noted that these models do not yet fully capture the concept of behavioural swarms, as introduced in \cite{[BHO20]}, where behaviour is modelled through an explicit activity variable that plays the role of an independent dynamical variable rather than a fixed parameter. This distinction is central to the development of a mathematical framework for living systems.

\vskip0.2cm
\noindent$\bullet$ \textbf{Hierarchical swarms:} Hierarchical structures in which particles have different roles have been explored in \cite{[S2008]}. In these swarms, lower-ranking particles follow the actions of higher-ranking particles. This type of organisation can be observed in animal groups (e.g.\ birds and fish) as well as human systems (e.g.\ the military, organisations, and transportation). The Cucker--Smale model can incorporate hierarchical leadership by limiting the summation in the alignment equation of the original model \(\eqref{eq:cucker-smale-model}\) to a specific group, known as the leader set of particle \(i\). This leader set comprises all the particles that influence particle \(i\) and has been discussed in works such as \cite{[CH08],[Li14],[LH15]}.

\vskip0.2cm
\noindent$\bullet$ \textbf{Models with herding:} This variant, proposed in \cite{[BCLY19]}, captures herding behaviour in financial markets by interpreting asset prices as positions and preferences as velocity within a modified CS model. The dynamics depend on particles' assessments, asset favourability and market-signal perturbations. Here, asset favourability represents a particle's tendency to increase its valuation of an asset, acting as a ‘velocity’ that drives the evolution of its subjective price. These dynamics are influenced by interactions with the prices and preferences of other particles, as well as differences between the particle's view and external market signals. Unlike traditional financial models, which focus on sequential analysis, this approach incorporates collective dynamics. It also differs from standard flocking models in that it allows particles to share positions in financial space. A kinetic-level extension was later derived in \cite{[BCKY19]} to study variations in subjective asset prices and favourability around their mean values.

\vskip0.2cm
\noindent$\bullet$ \textbf{Thermodynamic CS model:} Inspired by multi-temperature fluid mixtures, the thermodynamic CS model introduced in \cite{[HR17]} expands the CS framework by adding an internal variable representing the temperature of the particles. In this model, alongside the equations governing position and alignment dynamics, an additional equation is included to describe how the particle temperature evolves over time. Specifically, the temperature dynamics are governed by a nonlinear ordinary differential equation that reflects energy exchange between particles based on differences in inverse temperatures.

\vskip0.2cm
\noindent$\bullet$ \textbf{Stochastic swarms:} The Cucker--Smale models discussed thus far do not account for interactions between particles and their environment. As Ha et al.\ proposed in \cite{[HLL09]}, one approach to incorporating such interactions is to introduce stochastic perturbations into the deterministic dynamics. This results in the Stochastic Cucker--Smale (SCS) model, in which noise terms are added to the velocity equations. These noise terms are usually modelled as independent and identically distributed (\emph{i.i.d.}) \(d\)-dimensional Wiener processes characterised by mean zero and a covariance structure (i.e.\ white noise). This stochastic noise captures the random effects of the environment, resulting in a system described by stochastic differential equations. White noise therefore models external perturbations and enables the study of asymptotic flocking behaviour in the presence of randomness.

\subsection{Towards the complexity of living swarms} \label{sec:2.3}

The main properties of the CS model have been described in previous subsections, alongside a review of models applied to various real-world systems. In particular, applications to the dynamics of financial markets and thermodynamic multi-agent systems were considered. We now turn to the search for a unified theory of living systems (see~\cite{[CLE19]}), inspired by the philosophy proposed in Nowak's seminal book~\cite{[Nowak]} (see also~\cite{[NM01]}), and to the conceptual difficulties of dealing with living systems~\cite{[MAY],[REED]}. The goal is to understand how the modelling of deterministic swarms has evolved to address the modelling of the collective behaviour of living—i.e., complex-systems.

The reference for this topic is the book~\cite{[BBGO17]}, in which the authors, also motivated by applications, suggested that mathematical models of living systems should be derived from differential structures capable of incorporating the complexity features of living systems as far as possible. Indeed, living systems have the ability to develop specific strategies, which we can call `behavioural ability', to promote their own well-being. In particular, the following aspects are taken into account:

\vskip.2cm \noindent i) The ability to express behavioural actions that affect mechanics, and vice versa.

\vskip.2cm \noindent ii) The heterogeneity in their expression.

\vskip.2cm \noindent iii) Aggregation into groups expressing the same function.

\vskip.2cm The key idea is that a swarm has organisational abilities, usually referred to as `swarm intelligence', which control the collective dynamics. Further information can be found in references \cite{[BW1989],[KKHP]}. This interpretation can also be applied to particle methods (see, for example, Reia et al.~\cite{[RAF19]}). It has been proposed in~\cite{[BDL24]} that the idea of deriving a differential system suitable for describing swarm intelligence can be further developed to support the theoretical framework of artificial intelligence. The next section will report on and critically analyse some papers that have made important contributions consistent with the above features within the general framework of swarm theory.

An interesting example of the theoretical and empirical study of self-organisation was developed in \cite{[Ballerini]} through a physico-mathematical theory of topological interactions, which is based on the concept that each entity in a swarm interacts with a fixed number of other entities. This paper goes far beyond technical assumptions, investigating the dynamics of swarms in relation to their self-organisation. Additional studies have built on this seminal paper, motivated in part by the quest to understand the dynamics of self-organisation in living systems (see, for example, Cavagna et al.~\cite{[CavagnaA],[CavagnaB]}).

These studies were followed by a mathematical interpretation (see \cite{[BS12]}), in which the sensitivity domain was calculated using a differential system derived from the physical theory proposed in \cite{[Ballerini]}. This sensitivity domain was related to the visibility domain, which can be modified by external actions, such as an attack by a predator, or by natural heterogeneity within the swarm. 

\begin{figure}[htb!]
    \centering
    \includegraphics[width=0.7\textwidth]{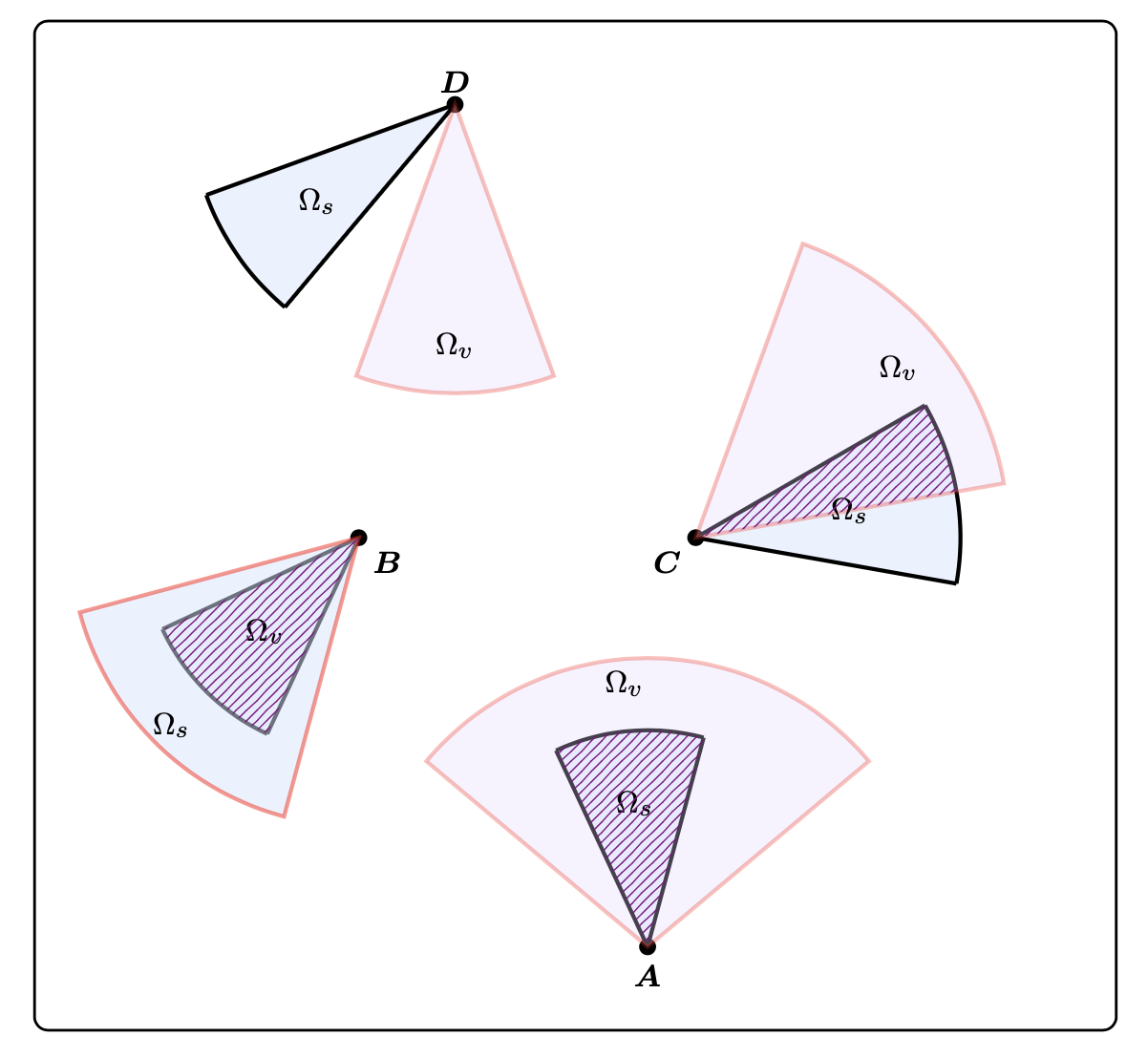}
    \caption{\textbf{Characterisation of visibility and sensitivity interaction domains.} 
A schematic representation of the geometric relationship between the visibility domain $\Omega_v$ (light red) and the sensitivity domain $\Omega_s$ (light blue) for four representative agents (A, B, C, D). Each case illustrates a distinct topological configuration: 
\textbf{(A)} For agent A, the condition $\Omega_s \subset \Omega_v$ is depicted, where the sensitivity domain is fully contained within the perceptual range. 
\textbf{(B)} Agent B illustrates the case $\Omega_v \subset \Omega_s$, where the perceptual range is more restrictive, imposing a physical constraint on potential interactions. 
\textbf{(C)} Agent C shows a partial intersection, $\Omega_v \cap \Omega_s \neq \emptyset$, where only a subset of the sensitivity domain is perceivable. 
\textbf{(D)} Finally, agent D represents the case of disjoint domains, $\Omega_v \cap \Omega_s = \emptyset$, resulting in a null interaction set.}
    \label{fig:vis_sens_domains}
\end{figure}

Figure~\ref{fig:vis_sens_domains} illustrates four scenarios involving a-particles (A, B, C and D), each of which moves with a velocity $\bv$ that defines its visibility domain $\Omega_v$ (light red). Its sensitivity domain $\Omega_s$ (light blue) depends on the local density of surrounding particles.

\begin{itemize}[itemsep=2mm]
\item Particle A represents the ideal case, where the visibility domain fully contains the sensitivity domain. In this configuration, the particle can access all the information necessary for interaction, ensuring an optimal response to its environment.

\item Particle B illustrates the opposite situation, where the sensitivity domain is larger than the visibility domain. Consequently, a part of the sensory region is outside the visual field, resulting in incomplete perception and potential loss of relevant interaction data.

\item Particle C shows a configuration where the visibility and sensitivity domains partially overlap, but neither is entirely contained within the other. This partial overlap leads to a partial loss of information, which may affect the particle's ability to respond accurately to stimuli within the interaction range.

\item Particle D presents the most critical scenario where the visibility domain $\Omega_v$ and the sensitivity domain $\Omega_s$ do not overlap at all. In this case, the particle is completely blind to its sensory field, resulting in a total loss of information.
\end{itemize}

However, the dynamics of the activity variable were not interpreted in~\cite{[BS12]}. This paper therefore serves as a bridge to the theory of behavioural swarms, which is reviewed and critically examined in the next section.

\section{The mathematical theory of behavioural swarms}\label{sec:3}

Traditionally, the study of swarms of classical particles has focused on mechanical and kinematic aspects, often treating the particles as simple, self-propelled entities governed by velocity alignment and attraction–repulsion forces, as well as environmental constraints. This literature is reviewed in Section~2. In contrast, this paper is devoted to the mathematical modelling of living systems. Therefore, we have developed a mathematical theory of behavioural swarms to take into account the specificity of living matter. In particular, we consider swarms composed of entities called active particles, or a-particles for short. These replace classical particles.

The difference between classical and a-particles is that the microscopic state includes an activity variable, which models their specific behavioural ability. As discussed in Section~2, these concepts are important for modelling living entities. This section outlines the key features of the theory and its potential applications, which will be discussed in the next section. This framework captures the interplay between individual activity levels and collective motion patterns, with mechanical interactions between particles affecting behavioural states. This formulation gives rise to a class of coupled differential systems in which the evolution of internal states influences motion, and vice versa. Paper~\cite{[BHO20]} serves as the main reference for this section.

The presentation will include a discussion of the relationship between our proposed framework and classical swarm models. We will highlight its generality and potential applications.

This section is organised as follows. In Section~\ref{subsec:3.1}, we develop the fundamental equations governing behavioural swarms, highlighting key features such as non-local interactions and hierarchical decision-making. We conclude the subsection by discussing the relationship between our proposed framework and classical swarm models, highlighting its generality and potential applications. In Section~\ref{subsec:3.2}, we illustrate how the mathematical framework of behavioural swarms, introduced in Section~\ref{subsec:3.1}, can be applied to model two representative scenarios drawn from behavioural economics~\cite{[KST20]}, crowd dynamics~\cite{[BHLY24],[KLY25]}, and vehicular traffic \cite{[BDF17],[CDF07],[Zagour23]}. Then, in Section~\ref{subsec:3.3}, we provide a brief review and critical analysis of the literature on the applications of behavioural swarm theory to the modelling and simulation of real-world systems.

\subsection{General structures}\label{subsec:3.1}

The modelling of behavioural swarms is concerned with a class of dynamical systems proposed to describe the motion of agents whose interactions are governed by both mechanical and socio-behavioural rules. This paradigm represents a departure from classical particle models, which often treat agents as homogeneous automatons. By contrast, the approach presented herein endows each agent with the capacity to develop adaptive strategies, leading to rich behavioural dynamics that can dictate its collective motion. The mathematical theory under review aims to provide a formal framework for exploring how this intricate interplay between internal states and physical dynamics might shape a swarm's evolution.

At its core, the theory postulates an enrichment of the microscopic state space. The state of an individual particle $i$ is no longer defined solely by its position $\bm{x}_i \in \mathbb{R}^d$ and velocity $\bm{v}_i \in \mathbb{R}^d$, but is augmented by a scalar internal variable, the \textit{activity} $u_i$. For mathematical tractability and to facilitate analysis across different systems, these variables are typically rendered dimensionless. Positions are scaled by a characteristic length $\ell$, velocities by a maximum attainable speed, and the activity $u_i$ is, by definition, normalised to the interval $[0,1]$, where the extremes $u_i=0$ and $u_i=1$ denote, respectively, quiescent and maximally active states. The collective configuration of the $N$-particle swarm is thus fully described by the concatenated state vectors $\bm{x} = (\bm{x}_1, \dots, \bm{x}_N)$, $\bm{v} = (\bm{v}_1, \dots, \bm{v}_N)$, and $\bm{u} = (u_1, \dots, u_N)$.

A consistent modelling framework for such systems must, as a necessary consequence of the complexity inherent to living entities, account for certain essential interaction mechanisms. Interactions within living systems, for instance, are rarely confined to immediate neighbours; they are often \textit{non-local}, mediated by long-range chemical or visual signals. Consequently, an agent's state can be influenced by distant particles, not just those in close proximity. Furthermore, these interactions are seldom linear. They depend not only on the microscopic states of the interacting pair but also on the emergent macroscopic properties of the swarm, such as its local density or overall polarisation. This interplay between scales gives rise to a critical feedback loop wherein the collective behaviour modulates individual dynamics, which in turn shapes the future evolution of the swarm.

The mathematical formalisation of these interactions hinges upon the concept of the \textit{interaction domain}, denoted $\Omega_i$, which can be taken as the intersection of the visibility domain and the sensitivity domain as illustrated conceptually in Figure~\ref{fig:vis_sens_domains}. This domain represents the specific subset of the swarm to which particle $i$ is actively paying attention at a given instant, and it is through this domain that non-local effects are realised. Whilst its precise geometry can vary—often stylised in $\mathbb{R}^2$ as a {\it ``cone''} of angle $\theta$ and maximum radius $R$—its key feature is its potential for dynamic adaptation. The domain $\Omega_i$ need not be static; it can expand or contract based on environmental cues or, more importantly, as a function of the agent's own activity level~$u_i$, allowing for a behavioural modulation of perception.

Finally, the strength of the coupling between any two particles, $i$ and $j \in \Omega_i$, is determined by an \textit{interaction rate}, $\eta_{ij}$. This rate is generally not a constant but a function of the particles' relative states, $\eta_{ij} = \eta(\bm{x}_i, \bm{v}_i, u_i; \bm{x}_j, \bm{v}_j, u_j)$, capturing the heterogeneous intensity of their interaction. While simplified scenarios may assume a constant $\eta_{ij}$, a more realistic model must account for its functional dependencies to capture the full spectrum of sophisticated behavioural responses observed in nature.

Interactions within an agent's domain, $\Omega_i$, are hypothesised to have a dual nature. These interactions include both a behavioural component, which governs the evolution of the agent's internal state and activity $u_i$, and a mechanical component that determines its physical motion. A key principle of this architecture is the \textit{decision-making hierarchy}. This principle enforces a clear temporal separation between the two components by defining a precise sequence of events: (i) the agent's internal state is updated, and (ii) this new behavioural state then informs and influences the agent's physical actions. This is not an arbitrary technical choice but a fundamental design feature, aimed at replicating the causal flow of living systems. In such systems, perception and internal changes necessarily precede and direct action. The hierarchy thereby creates a vital feedback loop: behavioural dynamics drive mechanical dynamics. The resulting physical changes, in turn, modify the agent's environmental perception, thereby affecting its subsequent behavioural evolution.

These principles are formalised within the following system of coupled, second-order ordinary differential equations, which describes the evolution of particle $i$'s state variables:
\begin{equation}\label{eq:swarm-general-structure}
\begin{cases}
\dfrac{d u_i}{dt} = z_i, \\[0.4cm]
\dfrac{d z_i}{dt} = \sum_{j \in \Omega_i} \eta_{ij} \cdot \chi_{ij}, \\[0.4cm]
\dfrac{d \bm{x}_i}{dt} = \bm{v}_i, \\[0.4cm]
\dfrac{d \bm{v}_i}{dt} = \sum_{j \in \Omega_i} \eta_{ij} \cdot \psi_{ij},
\end{cases}
\end{equation}
where the summation is performed over all particles $j$ within the interaction domain $\Omega_i$ of particle $i$.

Each term in this system has a precise physical and behavioural interpretation. In addition to the interaction rate $\eta_{ij}$ already introduced, the function $\chi_{ij}$ is the \textit{behavioural `force'} that drives the change in activity, representing social influences such as imitation or learning. Finally, the function $\psi_{ij}$ is the \textit{mechanical force}, which encompasses classical interactions like alignment, attraction and repulsion. Crucially, all three functions depend on the full state of the interacting pair, that is:
\begin{align*}
    \eta_{ij} &= \eta(\bm{x}_i, \bm{v}_i, u_i; \bm{x}_j, \bm{v}_j, u_j), \\
    \chi_{ij} &= \chi(\bm{x}_i, \bm{v}_i, u_i; \bm{x}_j, \bm{v}_j, u_j), \\
    \psi_{ij} &= \psi(\bm{x}_i, \bm{v}_i, u_i; \bm{x}_j, \bm{v}_j, u_j).
\end{align*}
This formulation ensures that the interactions are context-dependent and are modulated by the activity levels of the agents involved, thereby capturing the essence of a behavioural swarm.

It is important to elucidate the nature of the mechanical interaction term, $\psi_{ij}$. In its most comprehensive form, $\psi_{ij}$ can represent any form of mechanical influence, including non-conservative forces such as friction or hydrodynamics, which are prevalent in biological systems. Nevertheless, in numerous systems of physical interest where interactions are conservative, such as electrostatic, gravitational-like, or spring-like forces, this term can be derived from a pairwise interaction potential, $U(\bm{x}_i, \bm{x}_j)$. In such instances, the mechanical force is expressed as the negative gradient of the potential concerning the particle's position:
\begin{equation}
    \psi_{ij}(\bm{x}_i, \bm{x}_j) = -\nabla_{\bm{x}_i} U(\bm{x}_i, \bm{x}_j).
\label{eq:potential_force_v3}
\end{equation}
This formulation is not merely a formal simplification; it provides a robust connection to the methodologies of classical mechanics and stability theory. It facilitates the development of a global energy functional or a  Lyapunov function for the system, thereby enabling a meticulous analysis of equilibrium states and pattern formation through principles of energy minimisation. Recognising this relationship illustrates that the BST framework can inherently include gradient flow dynamics as a particular case, offering a direct approach to examine the convergence and stability of collective structures within an energy landscape---a perspective comprehensively developed within the broader context of interacting multi-agent systems~\cite{[PT13]}.

The true power of the general structure~\eqref{eq:swarm-general-structure} lies in its adaptability. It can be specialised to model a remarkable variety of phenomena by focusing on different components of the system. We highlight two important conceptual specialisations that illustrate this flexibility.

\subsubsection*{Specialisation 1: Spatially homogeneous systems (dynamics of activity only)}

In many systems of interest, particularly in the social sciences, the spatial arrangement of agents is secondary to the dynamics of their internal states. Consider, for instance, models of opinion formation or the collective evolution of sentiment in financial markets. In such scenarios, one can make the simplifying assumption that the system is \textit{spatially homogeneous}, effectively disregarding the kinematic equations for position and velocity. The framework then reduces to a system governing only the behavioural dynamics:
\begin{equation}\label{eq:activity-only}
\begin{cases}
\dfrac{d u_i}{dt} = z_i, \\[0.4cm]
\dfrac{d z_i}{dt} = \sum_{j=1,\, j\neq i}^{N} \eta_{ij}(u_i,u_j) \cdot \chi_{ij}(u_i,u_j),
\end{cases}
\end{equation}
where the interaction domain $\Omega_i$ is typically assumed to encompass the entire population. In this case, the `space' of interaction is abstract rather than physical and is defined by the features of the activity vector~$\bm{u}$. The interaction rate $\eta_{ij}$ and the behavioural force $\chi_{ij}$ now depend only on the relative activity levels between agents. This focused approach allows for detailed analytical and numerical studies of how behavioural states evolve, synchronise or polarise within a population, driven purely by non-physical social influence.

\subsubsection*{Specialisation 2: Swarms on networks}

The framework also offers a more structured treatment of space by replacing the continuous Euclidean setting with a discrete network topology. This is particularly relevant for modelling systems where interactions are constrained by a predefined architecture, such as social networks, epidemiological contact networks, or neural circuits. In this specialisation, agents occupy the nodes of a graph $\mathcal{G} = (\mathcal{V}, \mathcal{E})$, where $\mathcal{V}$ is the set of vertices (agents) and $\mathcal{E}$ is the set of edges (connections). The continuous position vector $\bm{x}_i$ loses its meaning, and the sensitivity domain $\Omega_i$ is naturally defined as the set of nodes adjacent to node $i$, i.e.\ $\Omega_i = \{j \in \mathcal{V}\mid(i,j) \in \mathcal{E}\}$. The governing equations for an agent $i \in \mathcal{V}$ become:
\begin{equation}\label{eq:network-swarm}
\begin{cases}
\dfrac{d u_i}{dt} = z_i, \\[0.4cm]
\dfrac{d z_i}{dt} = \sum_{j \in \Omega_i} \eta_{ij} \cdot \chi_{ij}, \\[0.4cm]
\dfrac{d \bm{v}_i}{dt} = \sum_{j \in \Omega_i} \eta_{ij} \cdot \psi_{ij},
\end{cases}
\end{equation}
where the state variable $\bm{v}_i$ is no longer a physical velocity but can be reinterpreted as another dynamic attribute of the node, such as an opinion vector~\cite{[HK02],[MT11B]}. In this abstract, topological setting, all state variables are re-contextualised from kinematic quantities to descriptors of internal, non-physical attributes. The activity $u_i$, for instance, sheds its role as a modulator of physical speed and is elevated to represent the \emph{intensity} or \emph{conviction} associated with the opinion state $\bm{v}_i$. A high $u_i$ may signify a zealous advocate, an agent highly committed to its opinion and thus more influential in its interactions, whereas a low $u_i$ could denote an undecided agent, more susceptible to the influence of its peers. The derivative, $z_i = du_i/dt$, thus acquires the meaning of \emph{behavioural momentum}---the rate at which an agent's conviction is changing. The interaction functions now depend on the states of connected nodes, and the rate $\eta_{ij}$ can be directly modulated by the weight of the edge $(i,j)$, representing the strength of the social tie. This specialisation provides a rigorous way to couple agent-based behavioural dynamics with complex, fixed interaction topologies, bridging the gap between swarm modelling and the statistical physics of social dynamics~\cite{[CFL09]}.

Furthermore, the framework can easily be extended to describe \textit{heterogeneous swarms} composed of multiple distinct populations, or \textit{Functional Subsystems (FSs)}. This is essential for modelling systems with functionally diverse agents, such as predator–prey ensembles or buyer–seller markets. In such cases, interactions occur both within each subsystem (intra-FS) and between different subsystems (inter-FS), allowing for a far more detailed and realistic description of complex ecological or economic systems. The governing equations follow the same structure as~\eqref{eq:swarm-general-structure}, but with interaction functions tailored to the specific nature of each subsystem pairing.

A final structural remark concerns a fundamental conservation property. The framework as presented is \textit{number-conservative}, meaning the total number of particles $N$ remains constant over time. This is a crucial property for ensuring physical consistency in many models, as it precludes the spontaneous creation or destruction of agents.

A crucial test of any generalised theory is its ability to recover established classical models within a well-defined limit. The Cucker–Smale (CS) model, as discussed in Section~2, can be seen as emerging from our framework by imposing a set of strong constraints that effectively collapse the behavioural dimension of the state space. First, one assumes the behavioural dynamics are trivial, rendering the activity $u_i$ a constant parameter rather than a dynamic variable. Second, the interaction mechanisms are simplified: the rate $\eta_{ij}$ is assumed constant, and the mechanical interaction $\psi_{ij}$ is specified as a pure velocity-alignment term, dependent only on the particles' positions and velocities:
\[
\psi_{ij}(\bm{x}_i, \bm{v}_i; \bm{x}_j, \bm{v}_j) = a_{ij}(\|\bm{x}_i-\bm{x}_j\|)\,(\bm{v}_j-\bm{v}_i).
\]
Under these constraints, the rich structure of the BST framework reduces to the classical Cucker–Smale system, demonstrating that our proposed theory is a consistent and natural generalisation.

Conversely, the general nature of the structure~\eqref{eq:swarm-general-structure} allows it to encompass other, more advanced models as specific instances. For example, the thermodynamic CS model introduced in~\cite{[HR17]} can be recovered within our framework. This is achieved by interpreting the activity variable $u_i$ as a measure of particle ‘temperature’ and selecting appropriate functional forms for the behavioural and mechanical interaction terms, $\chi_{ij}$ and $\psi_{ij}$, that model energy exchange. The thermodynamic model then emerges as a particular case. This shows that the BST framework is not merely an extension, but is proposed as a \textit{unifying mathematical structure} capable of describing a spectrum of complex agent-based systems under a single, coherent formalism.

\subsection{Applications to behavioural economics and crowd dynamics}\label{subsec:3.2}
This section demonstrates the application of the mathematical framework of behavioural swarms discussed in Section \ref{subsec:3.1}. As selected examples, we consider two specific case studies related to behavioural economics and crowd dynamics. Initial insights into the use of behavioural swarms for behavioural economics modelling can be found in \cite{[BBST24]}. Further development is presented in \cite{[KST20]}. For a comprehensive review of agent-based methods, see \cite{[BMST15]}. Recent advances in crowd dynamics are discussed in~\cite{[BHLY24],[KLY25]}.

\subsubsection{Price dynamics in a decentralised market}
A modelling approach consistent with the principles of our framework is presented in \cite{[KST20]}. This study builds on the concept of cherry-picking dynamics. It emphasises how buyers’ evaluation criteria, balancing price and perceived quality of goods, shape collective market behaviour. Building on earlier work (see \cite{[BBST24]}), it provides a more refined perspective on price evolution. The general structure of the model discussed here follows the formulation outlined in \cite{[KST20]}.

In an open market with $N$ sellers and $M$ buyers trading a given good, prices are determined by the interactions between these two subsystems. Each seller sets a price for their product. Each buyer sets a reservation price representing the maximum amount they are willing to pay. A key feature of this system is its asymmetry: while sellers make their prices public, buyers' reservation prices remain private. This asymmetry affects the way interactions take place and influences the market's overall dynamics.

The macroscopic influence of the market is represented by the mean price levels
\begin{equation}
\mathbb{E}[u] = \frac{1}{N} \sum_{s=1}^N u_s, \quad \hbox{and} \quad \mathbb{E}[w] = \frac{1}{M} \sum_{b=1}^M w_b,
\end{equation}
where these averages reflect the collective behaviour of sellers and buyers, impacting individual pricing strategies and adjustments.

The price evolution is governed, in the model, by the following system of equations
\begin{equation}
\begin{cases}
\frac{du_s}{dt} = v_s, \\[0.4cm]
\frac{dw_b}{dt} = z_b, \\[0.4cm]
\frac{d v_s}{dt} = \frac{1}{M} \sum_{q=1}^M \eta_{sq}(u_s,w_q) \psi_{sq}(u_s, w_q, v_s, z_q) + \mu_s(u_s, \mathbb{E}[u]) \varphi_{s}(u_s, \mathbb{E}[u]),   \\[0.4cm]
\frac{d z_b}{dt} = \frac{1}{N} \sum_{q=1}^N \eta_{bq}(w_b, u_q) \psi_{bq}(w_b,u_q,z_b,v_q) + \mu_b(w_b, \mathbb{E}[w]) \varphi_{b}(w_b, \mathbb{E}[w]).
\end{cases}
\end{equation}

According to the general framework developed in Section~\ref{subsec:3.1}:

\begin{itemize}[itemsep=2mm]
\item The function $\eta$ describes micro-micro interactions, i.e. direct negotiations between buyers and sellers. These interaction rates decrease as differences in price perception increase, reflecting asymmetry in the flow of information. The interaction rate from buyer to seller is given by
$$
\eta_{b,s} \approx \eta_s = \eta_0 \exp\left( - \frac{\rho}{\varepsilon} u_s \right),
$$
and from seller to buyer
$$
\eta_{sb} = \eta_0 \exp\left( - \frac{1}{\varepsilon}  \frac{|u_s - w_b|}{w_b} \right),
$$
where $\eta_0 > 0$ is the baseline interaction rate, $\rho \geq 0$ measures how slowly sellers are to change their prices, and $\varepsilon = \frac{N}{M}$ is the ratio of sellers to buyers. These expressions reflect two important features. Firstly, sellers tend to adjust prices more slowly (i.e., sticky prices). Secondly, buyers are less likely to interact if the seller's price differs greatly from their own estimate of value.

\item The function $\mu$ models micro-macro interactions, whereby individuals adjust their prices based on the overall population trend. Once again, the adjustment is asymmetric. Sellers adjust their prices at a constant rate
$$
\mu_s = \mu_0,
$$
where $\mu_0 > 0$ is the basic adjustment rate. Buyers, however, adjust their prices depending on how far they are from the average buyer valuation
$$
\mu_b = \mu_0 \exp\left( - \frac{1}{\varepsilon} \frac{|w_b -\mathbb{E}[u]|}{w_b} \right).
$$
This means that buyers react more when their valuation differs significantly from the market average. In contrast, sellers are less responsive to such differences and tend to persist with their pricing behaviour.

\item The function $\psi$ captures direct price-setting interactions, where one individual influences the other's price decision. When a buyer influences a seller
$$
\psi_{b,s} = \alpha u_s \operatorname{sign}(w_b - u_s),
$$
and when a seller influences a buyer
$$
\psi_{s,b} = \beta (u_s - w_b),
$$
where $\alpha$ and  $\beta$  are non-negative numbers that determine the strength of these reactions. These interactions resemble negotiations: if the buyer's valuation exceeds the seller's current price, the seller may increase the price, and vice versa.

\item Finally, $\varphi$  also incorporates adjustments based on market-wide expectations.  Agents tend to align their behaviour with the average price perceived within their group. For sellers, this adjustment is given by
$$
\varphi_s = \gamma (\mathbb{E}[u] - u_s),
$$
and for buyers by
$$
\varphi_b = \kappa (\mathbb{E}[w] - w_b),
$$
with $\gamma, \kappa \geq 0$ representing the tendency to follow the average market price.
\end{itemize}

Based on these interaction rules, \cite{[KST20]} develops two explicit models, referred to as Model 1 and Model 2.

Model 1 introduces buyer heterogeneity through fixed group strategies, whereby buyers are categorised based on their preferences (quality, price-quality ratio or price). This categorisation leads to coordinated seller selection and segmented market behaviour. In contrast, Model 2 features individualised buyer preferences via reservation qualities, which results in dynamic, personalised seller selection and more realistic, heterogeneous interactions. Although both models involve seller price adjustments based on buyer engagement, Model 2 captures more refined competition and shows less coordination among buyers.

Both models show pseudo-inertia. This means that price changes happen slowly over time instead of all at once. The model also draws from Hayek's idea of decentralised markets. It emphasises how interactions between individuals can influence larger market trends. This approach has been successfully applied in agent-based pricing models, like the ones in \cite{[MMT19]}.

\subsubsection{Behavioural crowd dynamics}
A modelling approach for pedestrian crowd dynamics that is consistent with our framework has been developed in \cite{[BHLY24],[KLY25]}.  These studies introduce leadership dynamics. They distinguish between leaders with predefined trajectories and followers whose activity levels are adjusted based on interactions. This section follows the general structure outlined in these references.

The movement of a pedestrian crowd involves individuals interacting with each other. These interactions alter their stress levels and influence their walking speed and direction. The crowd is modelled as a functional system. In this system, pedestrians are considered to be a-particles, with their stress levels representing activity that adapts based on the local environment.

The activity of an individual is influenced by their interactions with nearby pedestrians within their visibility range $\Omega_i$. The local mean activity is given by
\begin{equation}
\mathbb{E}[u | \Omega_i] = \frac{1}{\rho_i} \sum_{j \in \Omega_i} u_j,
\end{equation}
where $ \rho_i $ represents the local density, namely the number of individuals perceived within $ \Omega_i $.

The framework outlined in Section \ref{subsec:3.1} is applied here. The state of the pedestrian $ i $ is defined by its position $ \bm{x}_i $, velocity $ \bm{v}_i $, and activity $ u_i $, with second-order dynamics governing both the activity variable and the mechanical variables. The model is given by
\begin{equation}
\begin{cases}
\frac{du_i}{dt} = z_i,\\[0.4cm]
\frac{dz_i}{dt} = \eta(\rho_i) \left[(1-\varepsilon)(\mathbb{E}[u | \Omega_i] - u_i)+\varepsilon(\mathbb{E}[u | \Omega_i] - u_i)\right], \\[0.4cm]
\frac{d \bm{x}_i}{dt} = \bm{v}_i,\\[0.4cm]
\frac{d \bm{v}_i}{dt} = \gamma_i (s[\rho_i, u_i]\bm{w}_i - \bm{v}_i),
\end{cases}
\end{equation}
and where 
\begin{itemize}[itemsep=2mm]
\item $\eta$ is the interaction rate. Specifically, $\eta = \eta(\rho_i)$ increases with the local perceived density $\rho_i$. This reflects the idea that pedestrians are more influenced by neighbouring individuals when surrounded by more people. A typical form for such a dependence is exponential
\[
\eta(\rho_i) = \bar{\eta} e^{\lambda \rho_i},
\]
where $\bar{\eta}$ and $\lambda$ are empirical parameters.

\item The parameter $\varepsilon \in [0,1]$ is a small weighting factor. It models disagreement in opinion/activity dynamics. The parameter $\gamma_i$ is the relaxation rate, which typically depends on the perceived densities in the current and target directions.
\[
\gamma_i = \bar{\gamma} \exp\left\{ - u_i \left( \rho_{\omega_i} - \rho_{v_i} \right) \right\}.
\]
Here $\rho_{v_i}$ and $\rho_{\omega_i}$ represent the perceived densities in the current direction $v_i$ and the intended direction $\omega_i$, respectively. This form captures weaker reactions when the target direction is too crowded.

\item The term $s[\rho_i, u_i]$ defines the preferred walking speed of the $i$-th agent and incorporates both physical (density) and  behavioural (activity) inputs. A commonly used expression is
$$
s[\rho_i, u_i] = \frac{1 + u_i}{2} \, \phi[\rho_i],
$$
where $\phi[\rho]$ models the classical fundamental diagram relating pedestrian speed and density:
$$
\phi[\rho] =
\begin{cases}
1, & 0 \le \rho \le 1/5, \\[0.4cm]
\frac{125}{32}(1 - \rho)(\rho^2 - \frac{4}{5}\rho - \frac{1}{5}), & 1/5 < \rho \le 1, \\[0.4cm]
0, & \rho > 1.
\end{cases}
$$
This formulation reflects the slowing of pedestrians in denser environments, modulated by emotional state $u_i$.

\item Finally, $\bm{w}_i$ is the preferred walking direction. It is represented as a convex combination of a preferred direction $\bm{v}_T$ and a stream-following direction $\bm{v}^S_i$. The latter is derived from neighbouring velocities.
$$
\bm{w}_i = (1 - u_i) \bm{v}_T + u_i \bm{v}^S_i, \quad \text{where} \quad \bm{v}^S_i = \frac{\sum_{j \in \Omega_i} g_{ij} \bm{v}_j}{\left\| \sum_{j \in \Omega_i} g_{ij} \bm{v}_j \right\|}.
$$
The weights $g_{ij}$ decay with distance, typically chosen as:
$$
g_{ij} = \frac{\phi(\| x_i - x_j \|)}{\sum_{k \in \Omega_i} \phi(\| x_i - x_k \|)}, \quad \text{with} \quad \phi(y) = \frac{1}{(1 + y^2)^{\beta/2}}.
$$
This design strikes a balance between individual intention and social conformity within the swarm. $u_i$ controls the extent to which each agent follows the stream versus the target.
\end{itemize}

An extension of this model includes leadership dynamics within the crowd. Here, pedestrians are classified as either leaders or followers. Leaders have full knowledge of the environment and the optimal route, while followers adjust their activity level based on their interactions with others, learning to navigate the space. In this context, activity refers to the learning level of followers.

Simulations developed in \cite{[KLY25]} show the dynamics of leaders and followers. In particular, leaders follow predefined trajectories and keep their activity variable constant. Followers adjust their activity through interactions. This affects their walking speed and direction, causing them to follow the leaders' trajectories.

\subsection{On a critical review of applications}\label{subsec:3.3}

The previous subsections reviewed the mathematical theory of behavioural swarms  and its application to the study of real-world systems. 
This theory has been  proposed as a  natural progression from traditional swarm theory, enabling the incorporation of extra features of living systems into the dynamics of systems comprising living, interacting entities.

The main feature of this theory is that the state of each interacting entity (i.e. each a-particle) includes a behavioural variable, as well as mechanical variables. In particular, activity influences mechanical dynamics, which then in turn contribute to activity dynamics.

In some special cases, the dependent variable of the dynamical system is the activity alone. This occurs when spatial dynamics do not play an important role in collective dynamics. This occurs, for instance, in the spatially homogeneous case or when particles communicate independently of their spatial location.

The two case studies selected in the previous subsection are characterised by this difference.  Specifically, the first model is derived  from the broader framework of social dynamics and economics. By contrast, modelling crowd dynamics requires  an in-depth interpretation of  pedestrian social dynamics and their effect on spatial dynamics.

Before moving on to a critical analysis, it is important to  note that various parallel methods have been developed to model the collective dynamics of living systems. Rather than providing a detailed review, this report will offer a brief description of these methods and cite the minimum required bibliography.  However, this brief report will contribute to a critical analysis of the mathematical theory of behavioural swarms.
In particular, we mention the Kinetic Theory of Active Particles, KTAP for short, see~\cite{[BBGO17]}  and~\cite{[BBD21]}, and the Fokker–Planck–Boltzmann theory, FPB for short, see~\cite{[PT13]}. Both theories have been applied in various fields of applied science. 

The KTAP approach has been developed to model various complex systems, such as the immune competition between cancer and immune cells, see~\cite{[BD06]};  competition between immune cells and the SARS-CoV-2 virus, see~\cite{[BBO22],[BP21],[BK24]}; social dynamics and economics, see~\cite{[BE24],[DKLM17],[DL15],[DLO17]}, and several others as reviewed in~\cite{[BBGO17],[BBD21]}. The FPB theory has  primarily been used to model social dynamics and economics, see, for example, references~\cite{[FPTT17],[FPTT20]}, as well as several other areas  as reviewed in~\cite{[PT13]}. It has also been used to model the complex interaction between epidemics and social problems, see, for example~\cite{[B3EPT]}.

There are some technical differences between the KTAP and FPB theories and the BST approach under consideration. In fact, the derivation of the above kinetic theories requires the assumption that the number of a-particles is sufficiently large to support the assumption of continuity of the distribution functions. Conversely, the mathematical structures of the BST do not require this assumption. However, it suffers from computational complexity as the number of a-particles is high. Both the KTAP and BST approaches consider multiple interactions, which is very important when dealing with living systems. However, the KTAP approach accounts for both proliferative and destructive events,   whereas the BST approach does not.

All  of the above reasoning will be  reconsidered in Section 5,  where we will examine potential research perspectives aimed at  improving the descriptive capabilities of mathematical models of behavioural swarms.

\begin{figure}[htb!]
\includegraphics[width=\textwidth]{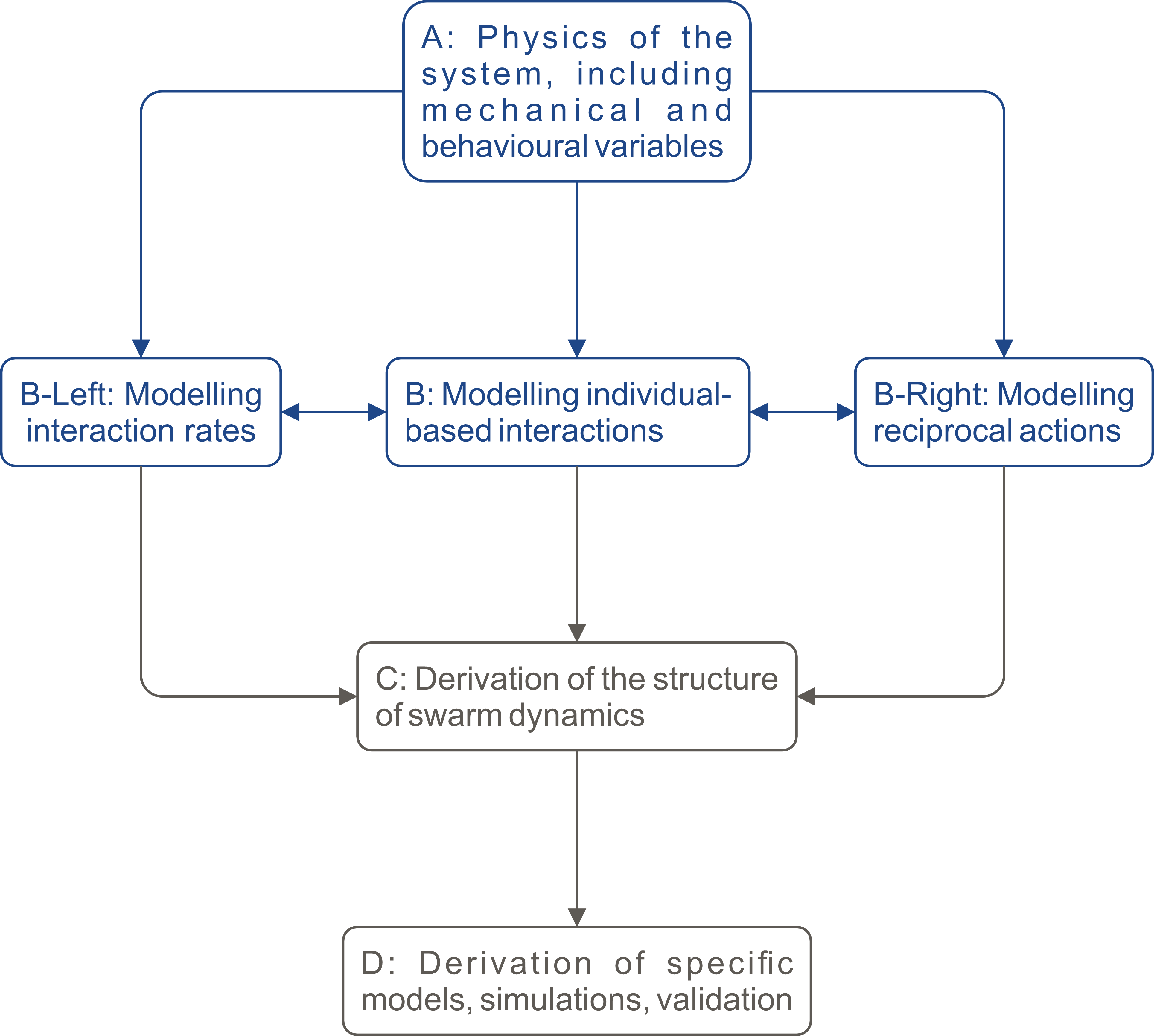}
\vskip-0.3cm
\caption{\textbf{Systematic methodology for behavioural swarm modelling.}
The flowchart outlines the hierarchical framework for model construction. The methodology is predicated on the definition of the system's core physics (\textbf{A}), which specifies the relevant mechanical and behavioural state variables. This foundation informs the modelling of three coupled aspects of agent interaction: the formulation of interaction rates (\textbf{B-Left}); the definition of individual-based interaction mechanisms (\textbf{B}); and the modelling of reciprocal actions (\textbf{B-Right}). The bidirectional arrows linking these components signify their mutual interdependence. The synthesis of these interaction rules yields the general mathematical structure for the swarm dynamics (\textbf{C}), from which specific models are derived for subsequent numerical simulation and empirical validation (\textbf{D}).}
\label{fig:modelling_methodology}
\end{figure}

\section{On the descriptive ability of behavioural swarms}\label{sec:4}

The previous sections have constructed the theoretical framework for behavioural swarms, emphasising their capacity to integrate activity variables and adaptive decision-making into collective dynamics. To validate and illustrate the descriptive power of this framework, this section transitions to numerical explorations.

Firstly, through targeted applications, we demonstrate how behavioural swarm models can generate rich dynamics. Specifically, we report two applications in Section~\ref{Subsec:4.1}: (i) the role of stress in shaping swarm behaviour \cite{[BHO20],[BHOY22]}, and (ii) modelling of the price in a market with cherry-picking behaviour \cite{[KST20]}. We begin with a brief description of the physical problem modelled in each case, followed by a summary of the mathematical structure and a concise presentation of the simulation results. These case studies demonstrate the framework's ability to capture emergent phenomena in socio-economic contexts, bridging theoretical constructions with observable dynamics.

Next, in Section~\ref{Subsec:4.2}, we investigate the influence of the {\it interaction domain} on panic propagation by selecting different visibility angles, and in Section~\ref{Subsec:4.3} we consider {\it multi-consensus} dynamics in behavioural swarms.

\subsection{A review of previous applications of behavioural swarms}
\label{Subsec:4.1}

\subsubsection{The role of stress in shaping swarm behaviour.}
Herein, we consider a class of models, introduced in~\cite{[BHO20]}, designed to explore collective motion wherein agent orientation is governed by a combination of an external signal—a global, preferred direction—and local interactions with neighbours. Central to this framework is the `activity' variable, a scalar quantity representing an internal state of stress or arousal that critically modulates the dynamics of alignment. The fundamental scientific question concerns how macroscopic, coordinated motion emerges from microscopic interaction rules. Mathematically, each particle's heading changes as a convex combination of its preferred direction and the average heading of its neighbours. This framework places the model within the broader context of consensus dynamics. We explore two distinct physical regimes: a first-order, overdamped model, which assumes instantaneous velocity adjustment, and a more physically realistic second-order model, which includes pseudo-inertia, accounting for rotational acceleration and damping forces that limit reorientation.

The first-order formulation, whose analytical properties for similar models were rigorously established in~\cite{[BHOY22]}, yields valuable insights across three canonical scenarios for the activity state. The numerical simulations in~\cite{[BHOY22]} indeed confirm the trends first observed in~\cite{[BHO20]}.

\textbf{Constant activity.} In this fundamental case, the activity is a prescribed, uniform parameter. It is observed that highly active particles, which assign more weight to the external directional signal, rapidly align and form highly coherent, polarised streams. Conversely, agents with low activity show a markedly slower relaxation towards the preferred heading, resulting in less synchronised movement. A particularly illuminating case occurs when the preferred direction rotates over time, for example, by tracing a circular path. The flocks adapt, with higher activity enabling a more responsive pursuit. This leads to tighter circular trajectories with smaller radii, a direct outcome of a reduced phase lag between the agent's heading and the time-varying external signal. This behaviour is strongly analogous to a damped, driven oscillator, where higher `activity' signifies a stronger coupling to the driving force.

\textbf{Dynamic activity based on consensus.} Here, we suggest that each agent's activity is itself a dynamic variable, evolving towards the average activity of its neighbours. This introduces a second, coupled consensus problem: the system aims for consensus not only in the kinematic heading angle but also in the internal behavioural state. This mutual adaptation mechanism fosters a gradual homogenisation of the swarm's responsiveness. The resulting behaviour is a global flocking state with noticeably improved cohesion compared to the constant activity scenario. By dynamically resolving internal differences, the swarm functions as a more unified entity, effectively filtering out the destabilising effects of individual heterogeneity.

\textbf{Asymmetric influence of high-activity neighbours.} This final scenario presents a non-reciprocal interaction, a feature of great importance in active-matter systems. Particles are affected solely by neighbours with higher activity levels. As a result, the most active agents act as emergent, mobile leaders, with their movement creating dynamic focal points around which the rest of the swarm organises. This clear information hierarchy, in which influence flows from the more active to the less active, prevents the `diluting' effect of less responsive agents on the consensus process. The outcome is a quicker and more reliable convergence towards structured, directional swarms than in the symmetric consensus case.

While the first-order formulation offers significant analytical tractability, it presupposes an idealised, instantaneous reorientation that neglects the physical realities of mass and momentum. We therefore extend our investigation to a second-order dynamical system, where inertia and damping introduce a finite timescale for changes in heading. The resulting trajectories are demonstrably smoother and less abrupt. Although herding phenomena still emerge, they do so over longer timescales, and the system can exhibit transient overshoots and oscillations absent in the overdamped limit. These results are of critical importance, highlighting the non-trivial role of inertial effects in systems where rapid changes in heading are physically constrained or energetically costly, as is the case in animal herds, dense crowds, or robotic swarms. The second-order model, therefore, provides a more faithful, albeit more complex, representation of many real-world systems, inspiring and motivating further research and development in these areas.


\subsubsection{Modelling price dynamics in a market with cherry-picking behaviour}

We now shift our focus to an application in mathematical economics, analysing a model of price dynamics in a market characterised by `cherry-picking' buyer behaviour, as introduced in~\cite{[KST20]}. This framework marks a significant departure from traditional general equilibrium models, which assume a centralised, Walrasian price-setting mechanism. Instead, the process of price discovery is endogenous, arising from the decentralised interactions of diverse agents within a well-defined market structure. The core of the model, explained in Section~3.2, comprises a population of buyers and sellers. The buyers, modelled as a group of utility-maximising agents, display `cherry-picking' behaviour: they actively seek out sellers offering the most favourable (i.e., lowest) prices. This creates a strong feedback loop between consumer choice and market structure, which we examine through two illustrative case studies.

\textbf{Fixed sellers' prices.} It is instructive to first consider a static market environment where sellers' prices are exogenously fixed. In this context, the buyers' search for the lowest price can be viewed mathematically as a consensus-seeking process on a static fitness landscape, where a seller's fitness is inversely proportional to their offered price. The model reliably predicts the formation of `macro-waves' of demand; these are, in essence, information cascades wherein the buyer population rapidly coordinates to form dense clusters around the most competitive sellers. This self-reinforcing dynamic results in a `winner-take-all' market structure, a form of extreme concentration. From an economic perspective, this outcome exhibits a limited and fragile form of efficiency. For the subset of buyers who successfully transact, the market is highly efficient, allocating them to the optimal available price. However, it also causes market failure for higher-priced sellers, who are entirely excluded from trade. The resulting allocation may be regarded as Pareto-efficient in the narrowest sense (no transacting buyer can be made better off), but this ignores the welfare of the excluded sellers and exposes the market to significant fragility should the `winning' seller fail.

\textbf{Adaptive sellers' prices.}  
A more sophisticated approach, more accurately reflecting real-world markets, endogenises the price-setting mechanism, enabling sellers to adjust their prices based on observed buyer demand.  
This converts the system into a co-evolutionary dynamic problem, where the price landscape and buyer distribution develop together.  
Sellers modify their prices using a simple, boundedly rational heuristic: a high density of local buyers encourages a price increase, while a lack of buyers prompts a price decrease.  
This feedback loop generates a complex dynamic reminiscent of Bertrand competition.  
At first, sellers with few buyers are forced to lower their prices, which attracts new waves of `cherry-picking' buyers.  
This often leads to a `race to the bottom', as sellers repeatedly undercut each other.  
Ultimately, the system self-organises towards a market-wide price consensus---a single, emergent price at which most transactions occur.  
The decentralised achievement of this market-clearing price is a significant result, showing how global order can develop from simple, local rules without any central oversight.  
This convergence improves allocative efficiency across the entire market.  
However, the transient dynamics can display considerable volatility before reaching this equilibrium, representing a period of fundamental economic uncertainty for market participants.  
The model therefore offers a rich, dynamic depiction of how competition and coordination interact to influence market outcomes.

 
\subsubsection{A synthesis of the applications: a theoretical dichotomy}

The applications presented here, while diverse, can be categorised into two profoundly different yet complementary theoretical classes. A thorough examination of these classes reveals the remarkable versatility of the behavioural swarm framework and highlights a fundamental conceptual split in its use: the distinction between models where dynamics occur in a physical space and those where dynamics are confined to an abstract space of internal states. This distinction depends on the ontological status of the activity variable, $u_i$.

The first, more traditional, category of applications is characterised by its clear integration within a physical or conceptual spatial domain. In these models, such as those examining stress-induced motion or market price formation, each agent has a state vector that includes both kinematic variables (e.g., position and direction) and the internal activity variable. Here, the activity $u_i$ functions as a crucial \textit{modulator} of behaviour. It is an endogenised parameter that governs how an agent perceives and responds to its environment---be it a global directional signal or a landscape of sellers' prices. When $u_i$ is treated as a dynamic quantity, as explored in recent theoretical extensions~\cite{[BHLY24],[KLY25]}, the model captures a feedback loop where an agent's internal state both influences and is influenced by its spatial interactions. This represents a significant paradigm shift from classical models, where agent properties are typically static, offering a powerful tool for studying adaptation and learning in spatially embedded systems.

The second class of applications signifies a more significant conceptual shift, where the dynamics occur not in physical space, but entirely within the abstract state space of the activity variables themselves. The model of cell differentiation, inspired by the work in~\cite{[Zhang]}, serves as the canonical example. In this formulation, there are no spatial coordinates; the agents do not `move' in the conventional sense. Instead, the activity variable $u_i$ ceases to be a mere modulator and becomes the \textit{locus of the dynamics itself}. The evolution of $u_i$ represents the entire behavioural change---for instance, a cell's progression along a differentiation pathway towards a specialised fate. The interactions are not mediated by spatial proximity. However, they are directly defined by the activity state values, leading to phenomena such as homophily-driven clustering or symmetry breaking. This class of models is therefore suitable for problems in opinion dynamics, social norm formation, and developmental biology, where the key question is not `where do agents go?' but `what do agents become?'.

Ultimately, the strength of the mathematical framework presented in this work lies in its ability to handle this dichotomy. By treating the activity variable as a primary component of the dynamical system, we develop a flexible theoretical structure that can unify the modelling of physical swarms and abstract differentiation processes. The key decision to include spatial variables influences which type of phenomenon is addressed, enabling a single, coherent mathematical language to describe a wide range of complex adaptive systems.

\subsection{Propagation of panic in Behavioural swarms with different interaction domains} \label{Subsec:4.2}

In this subsection, we consider the propagation of panic in a swarm of $N$ pedestrians, with states $(u_i, \bm{x}_i,\bm{v}_i)$, $i=1, \cdots, N$.  Here we take $u_i$ as the emotional state (panic level) for the $i$-th pedestrian, $(\bm{x}_i,\bm{v}_i)$ are the mechanical variables denoting their position and velocity. 
 
\begin{figure}[htb!]
\includegraphics[width=\textwidth]{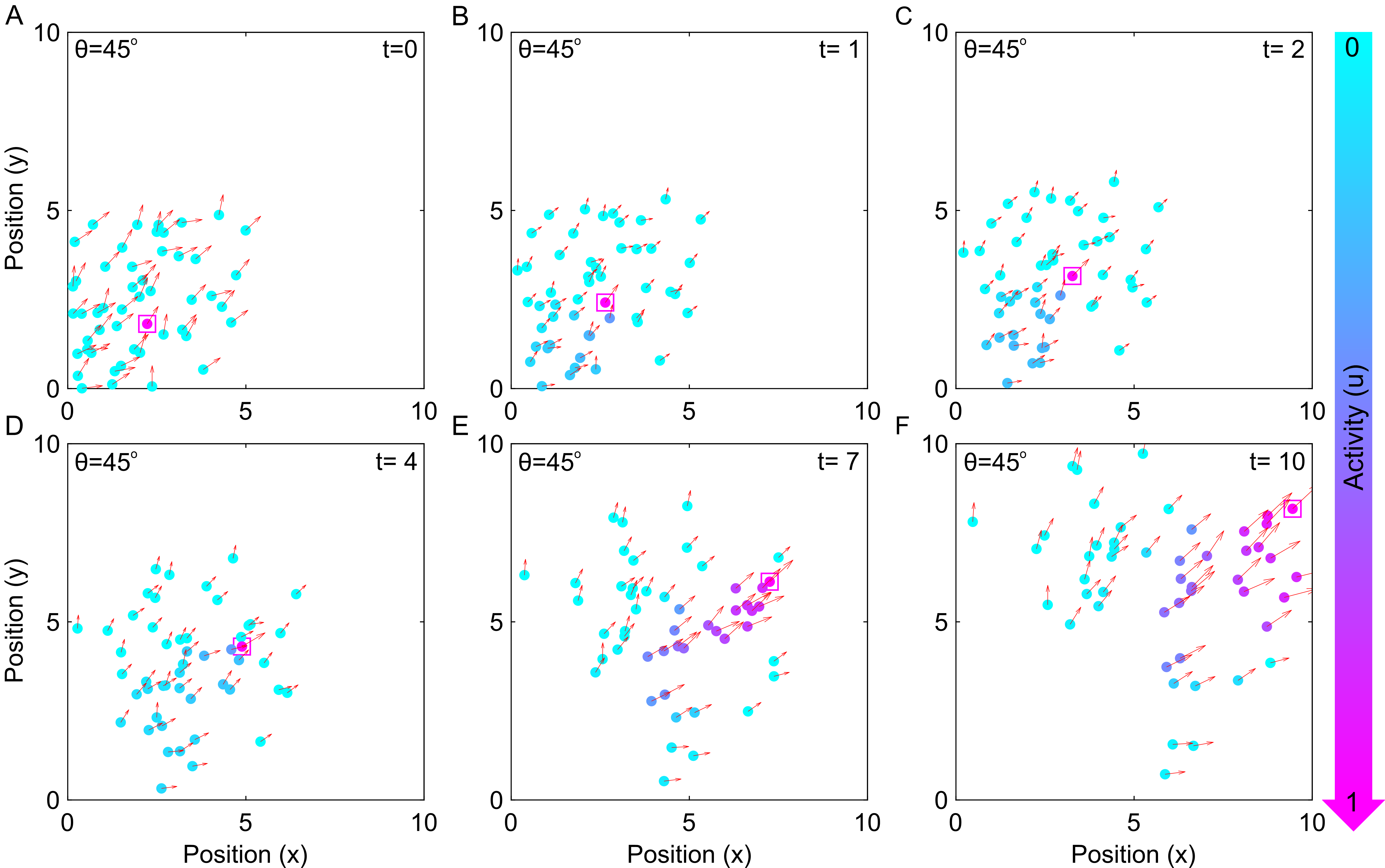}
\vskip-0.3cm
\caption{\textbf{Panic propagation with narrow visibility angle ($\boldsymbol{\Theta=45^{\circ}}$).}
Snapshots illustrating the dynamics of panic propagation for $N=50$ pedestrians governed by model \eqref{test1}. Agents are initially distributed in the domain $[0,10] \times [0,10]$. Each agent's state is represented by its position (dot), velocity vector (red arrow), and panic level $u_i \in [0,1]$ (colour scale: cyan calm, magenta panicked). The single, initially panicked agent is identified by a square marker. The panels (\textbf{A}-\textbf{F}) sequentially depict the gradual spread of the panic state from this source. Simulation parameters: visibility angle $\Theta=45^{\circ}$, visibility distance $R=5$, and relaxation rates $\beta_i = \gamma_i = 1$.}
\label{fig:panic_theta45}
\end{figure}

We assume that initially the pedestrians are randomly distributed in a square, for instance, $\Sigma_0 = [0,5]\times [0,5]$, each one walks with a random direction in the {\it north-east} sector as their {\it target direction} that 
 $$\bm{v}^T_i = (\cos \theta_i, \sin \theta_i), ~ \theta_i  \in (0,\pi/2), ~  i=1, \cdots, N.$$
The swarm moves freely without any panic, by taking $u_i=0$ for all pedestrians, except suddenly one pedestrian gets panicked, say, $u_{N}=1$.   
The initial configuration of the swarm at $t=0$ is shown in the initial configuration depicted in Fig.~\ref{fig:panic_theta45}(\textbf{A}), where the panicked pedestrian is highlighted with a small square.

 In the movement, each pedestrian has a {\it interaction domain} $\Omega_i = \Omega_i(\bm{x}_i,\bm{v}_i)$ in which he/she learns the emotional state of others and then adjusts his/her walking direction. 
 The interaction domain takes a sector that is symmetric with respect to the walking direction, with a visibility distance $R$ and visibility angle $\Theta$.

 The swarm can be described by the model    
\begin{equation}\label{test1}
\begin{cases}
\displaystyle  \frac{d u_i}{dt}  =   \beta_{i}  \sum_{j \in \Omega_i}  \psi(\| \bm{x}_j-\bm{x}_i\|) (u_j - u_i),  \\[0.4cm]
\displaystyle  \frac{d\bm{x}_{i}}{dt} =  \bm{v}_i,\\[0.4cm] 
\displaystyle  \frac{d\bm{v}_{i}}{d t} = \gamma_i  
  (s_i \, \boldsymbol{\omega}_i - \bm{v}_i    ),  
\end{cases}
\end{equation} 
where $ \beta_{i}$ and $\gamma_i$ are prescribed relaxation rates, which can be taken as 1 for simplicity, $s_i$ and $\boldsymbol{\omega}_i$ are the preferred speed and preferred walking direction of the $i$-th pedestrian, respectively, defined as 
 \begin{equation}\label{test2}
\begin{cases}
s_i = \frac{1+ u_i}{2}, \\[0.4cm]
\boldsymbol{\omega}_i   = (1 -u_i ) \bm{v}^T_i +  u_i \bm{v}_i^S,  \quad
  \bm{v}_i^S= \sum_{j \in \Omega_i}  \psi (\|\bm{x}_j - \bm{x}_i \|) \, \bm{v}_j .
  \end{cases}
\end{equation}
Here the communication function $\psi$ takes the form in \eqref{eq:cucker-smale-communication}. 
 
\begin{algorithm}[t!]
\KwIn{$N$, $h$, $T$, $\bm{v}^T_i$,  $\beta_i,\ \gamma_i$, $\psi$}
  \tcc{$N$ - number of a-particles;\\ 
  $h$ -  size of time step; \\
  $T$ - number of time steps; \\
  $\bm{v}^T_i$ - preferred direction; \\
  $\beta_i, \, \gamma_i$ - relaxation rates; \\
  $\psi$ -  communication weight function.}

 {\bf Initialization:} 
 
 ~$\quad$ Generate the initial states $(u_i, \ \bm{x}_i, \  \bm{v}_i)|_{t=0}$ for $i=1, \cdots, N$; 
 
  \For{$t = 0$ \textbf{to} $T-1$ }{ 
       \For{$i=1, \cdots, N$}{          
          Compute the interaction domain $\Omega_i[t] = \Omega_i(\bm{x}_i,\bm{v}_i)|_t$;  
       
             \For{$j\in \Omega_i[t] $}{
             Compute communication function $ \psi_{ij}[t] = \psi(\| \bm{x}_j[t]-\bm{x}_i[t]\|) $; 
             }  
       Update the activity:  
            $$  u_i[t+1] = u_i[t]+h\beta_i \sum\limits_{j \in \Omega_i[t]}  
               \psi_{ij}[t]   (u_j[t]-u_i[t]);$$\vskip-2mm  
       Compute  the preferred speed 
         $s_i[t+1]  = (1+ u_i[t+1] )/2 $; 
   
        Compute the preferred walking direction  
           $$\boldsymbol{\omega}_i [t+1]  = (1 -u_i [t+1]) \bm{v}^T_i +  u_i [t+1]\bm{v}_i^S, \  
                  \bm{v}_i^S= \sum_{j \in \Omega_i}  \psi_{ij}[t]   \bm{v}_j[t];$$\vskip-3mm 
                  
         Update the velocity: 
             $$ \bm{v}_i[t+1]=\bm{v}_i[t] + h\gamma_i 
             (s_i[t+1]   \boldsymbol{\omega}_i[t+1] - \bm{v}_i[t] ) ;$$\vskip-2mm 
             
         Update the position: 
            $$   \bm{x}_i[t+1]=\bm{x}_i[t]+h \bm{v}_i[t+1]; $$
    }
    }
    \KwOut{$u_i[t], \ \bm{x}_i[t], \ \bm{v}_i [t]$,  for $i = 1,\ldots N$, $t=1,\ldots T$.}
  \caption{Numerical simulation of the swarm model \eqref{test1}-\eqref{test2}.}
  \label{alg:1}
\end{algorithm}

We begin with a detailed numerical investigation designed to elucidate the dynamics of the panic model formulated in~\eqref{test1}--\eqref{test2}. Our primary objective is to explore the propagation of panic as an emergent, collective phenomenon and to characterise its dependence on the geometric constraints of agent interaction. The system of ordinary differential equations is solved using the computational scheme detailed in Algorithm~\ref{alg:1}.

Before proceeding to the specific numerical experiments, it is important to recall the foundational assumptions of the model. From the selection rule in~\eqref{test2}, we note that for a quiescent agent with an emotional state of $u_i=0$, the preferred walking direction $\boldsymbol{\omega}_i$ is identical to its current velocity vector, $\bm{v}_i$. Consequently, it is only a non-zero emotional state, $u_i > 0$, that induces a change in heading. The model also assumes a sparse swarm, a simplification that allows the exclusion of explicit collision-avoidance terms (such as short-range repulsive potentials) from the interaction dynamics. Furthermore, an agent's speed, $s_i = \|\bm{v}_i\|$, is defined as a monotonically increasing function of its emotional state alone, $s_i = f(u_i)$, thereby neglecting dependencies on local agent density. A more general treatment of these aspects can be found in~\cite{[KLY25]}.

To probe the influence of the interaction geometry on the swarm pattern, we perform two computational experiments with different visibility angles, $\Theta=45^{\circ}$ and $\Theta=90^{\circ}$, respectively, while holding the visibility distance fixed at $R=5$. This allows us to isolate the effect of the perceptual field of view on the propagation dynamics.

The first experiment, with a narrow visibility angle of $\Theta=45^{\circ}$, is depicted in the temporal snapshots of Fig.~\ref{fig:panic_theta45}. The system is initialised with a quiescent, spatially homogeneous swarm, into which a single `seed' of panic is introduced. One observes the formation of a remarkably coherent panic wave that propagates upwards through the domain. The underlying mechanism is a self-sustaining chain reaction: agents within the wave, having entered a state of high activity, exert a strong attractive influence on their forward, quiescent neighbours. As the visibility cone captures these neighbours, they are in turn rapidly recruited into the panic state, thus advancing the panic wave front.

\begin{figure}[htb!]
\includegraphics[width=\textwidth]{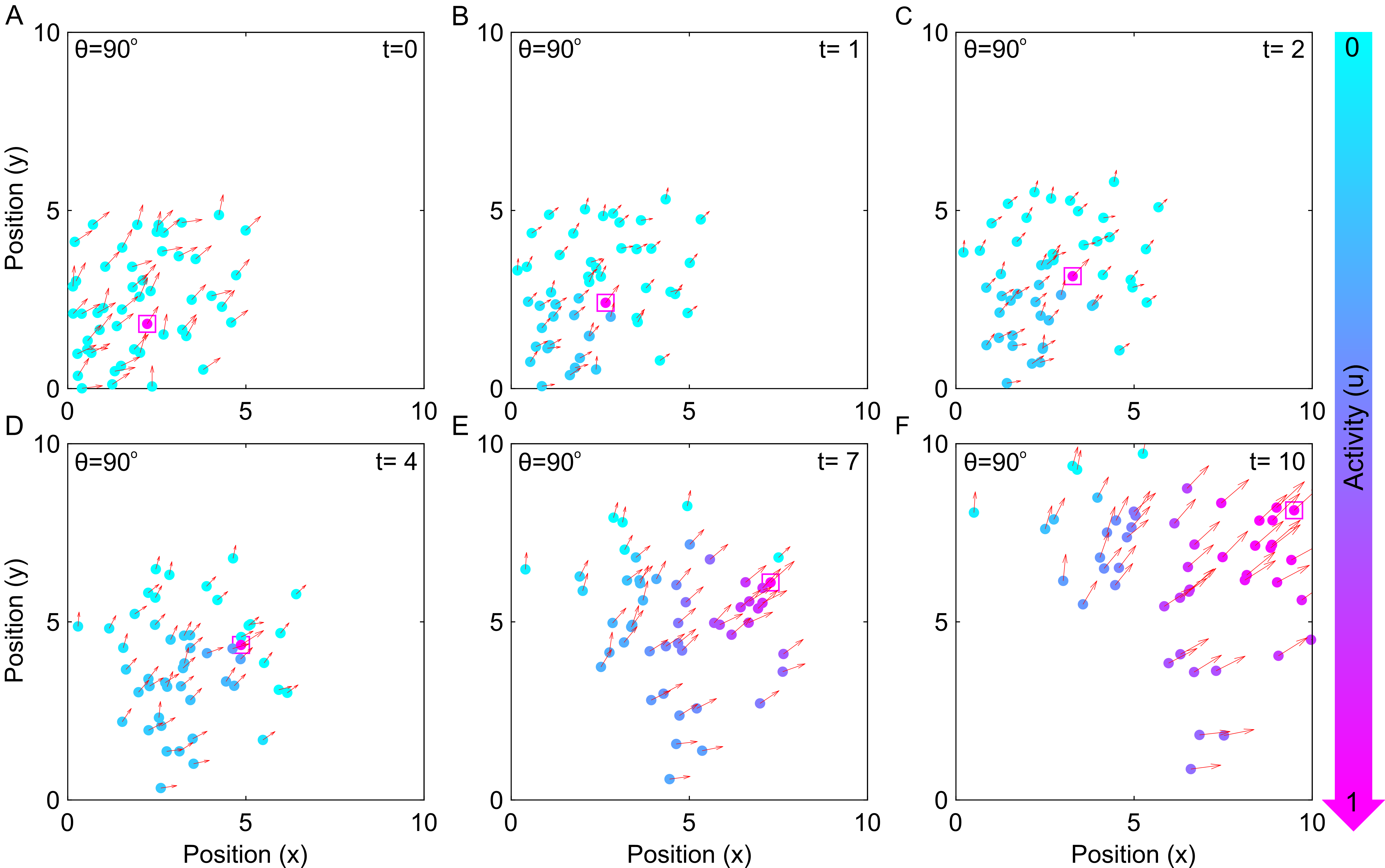}
   \vskip-0.3cm
\caption{\textbf{Enhanced panic propagation with a wide visibility angle ($\boldsymbol{\Theta=90^{\circ}}$).}
Dynamics of panic propagation for the same system presented in Fig.~\ref{fig:panic_theta45}, but with the visibility angle increased to $\Theta=90^{\circ}$. The simulation setup and visual conventions are identical. A direct comparison reveals that the wider field of view results in a visibly more rapid and extensive spread of the panic state. At the final time shown, $t=10$ (\textbf{F}), a significantly larger proportion of the swarm exhibits high panic levels ($u_i \to 1$) than in the narrow-angle scenario. All simulation parameters are identical to the previous case, with the exception of $\Theta=90^{\circ}$.}
\label{fig:panic_theta90}
\end{figure}

The second experiment, presented in Fig.~\ref{fig:panic_theta90}, employs a wider visibility angle of $\Theta=90^{\circ}$. A direct comparison with the previous case reveals two striking differences. Firstly, the panic wave propagates with a demonstrably greater velocity. This is a direct consequence of the enlarged perception cone, which allows the panic signal to be transmitted more effectively among agents, increasing the number of influential neighbours for any given agent. Secondly, the wavefront is significantly broader, indicating that the chain reaction is no longer confined to a narrow channel but can branch out, recruiting agents from a wider lateral cross-section. This directly links the microscopic parameter $\Theta$ to the macroscopic morphology of the emergent wave.

Furthermore, a key phenomenon revealed by these simulations is that the panic wave does not simply propagate; it \textit{accelerates}. Initially, when only a few agents are panicked, the emotional escalation progresses gradually. However, as the wave advances, a quiescent agent just ahead of the front is influenced by an increasingly large number of panicked agents within its perception cone. This enhances the weighted average forcing term in its dynamics, thereby reducing the transition time into the panic state. This non-linear amplification, clearly observable when comparing the progression in Fig.~\ref{fig:panic_theta45} and Fig.~\ref{fig:panic_theta90}, is a characteristic feature of the behavioural swarm model, demonstrating its ability to capture the essential physics of cascading failures and information spread in complex systems.

\subsection{Multi-consensus dynamics in behavioural swarms} \label{Subsec:4.3}

A slight modification to the model \eqref{test1} can show consensus and multi-consensus dynamics in behavioural swarms. More specifically, we assume that each a-particle in a swarm aligns its activity with others in the interaction domain, adjusts its heading angle (moving direction) to a desired one as the activity-weighted combination of target direction $\theta^\nu$ and stream direction $\theta^S$, and moves with a normalised speed $1$. Then we consider a consensus model
\begin{equation}\label{test3}
\begin{cases}
\frac{du_i}{dt}=  \beta_{i} \sum_{j \in \Omega_i } {\psi(\|\bm{x}_j -\bm{x}_i\|)}
(u_j - u_i ),  \\[0.4cm]
\frac{d\bm{x}_i}{dt}=(\cos{\theta_i},\sin{\theta_i}),\\[0.4cm]
\frac{d\theta_i}{dt} = \gamma_i (\bar\theta_i -\theta_i), \  
\bar\theta_i = u_i \theta^\nu + (1-u_i) \theta^S, \   
\theta^S =\sum_{j\in\Omega_i}{\phi(\|\bm{x}_j-\bm{x}_i\|)} \theta_j .
\end{cases}
\end{equation}

As noted in \cite{[BHO20],[BHKL25]}, the heading angle of each a-particle $\theta_i$ will tend to $\theta^\nu$ as $t\to \infty$, called {\it alignment} of the heading angle, under some assumptions on the initial states and the interaction domain. A higher activity level corresponds to quicker alignment of the heading angle to $\theta^\nu$.

To illustrate these dynamics, we perform a numerical simulation with $N=100$ agents under an all-to-all interaction topology ($\Omega_i = \{1,\dots,N\} \setminus \{i\}$ for all $i$). Agents are initialised with random positions in $[4,8]\times [4,8]$, headings in $[-\pi, \pi]$, and activities in $[0,1]$. We set the relaxation rates $\beta_i = \gamma_i =1$, the target angle $\theta^\nu = 0$, and employ communication functions $\psi(\cdot) \text{ and } \phi(\cdot)$ of the Cucker--Smale type, as specified in \eqref{eq:cucker-smale-communication}.

\begin{figure}[htb!]
  \centering
\includegraphics[width=\textwidth]{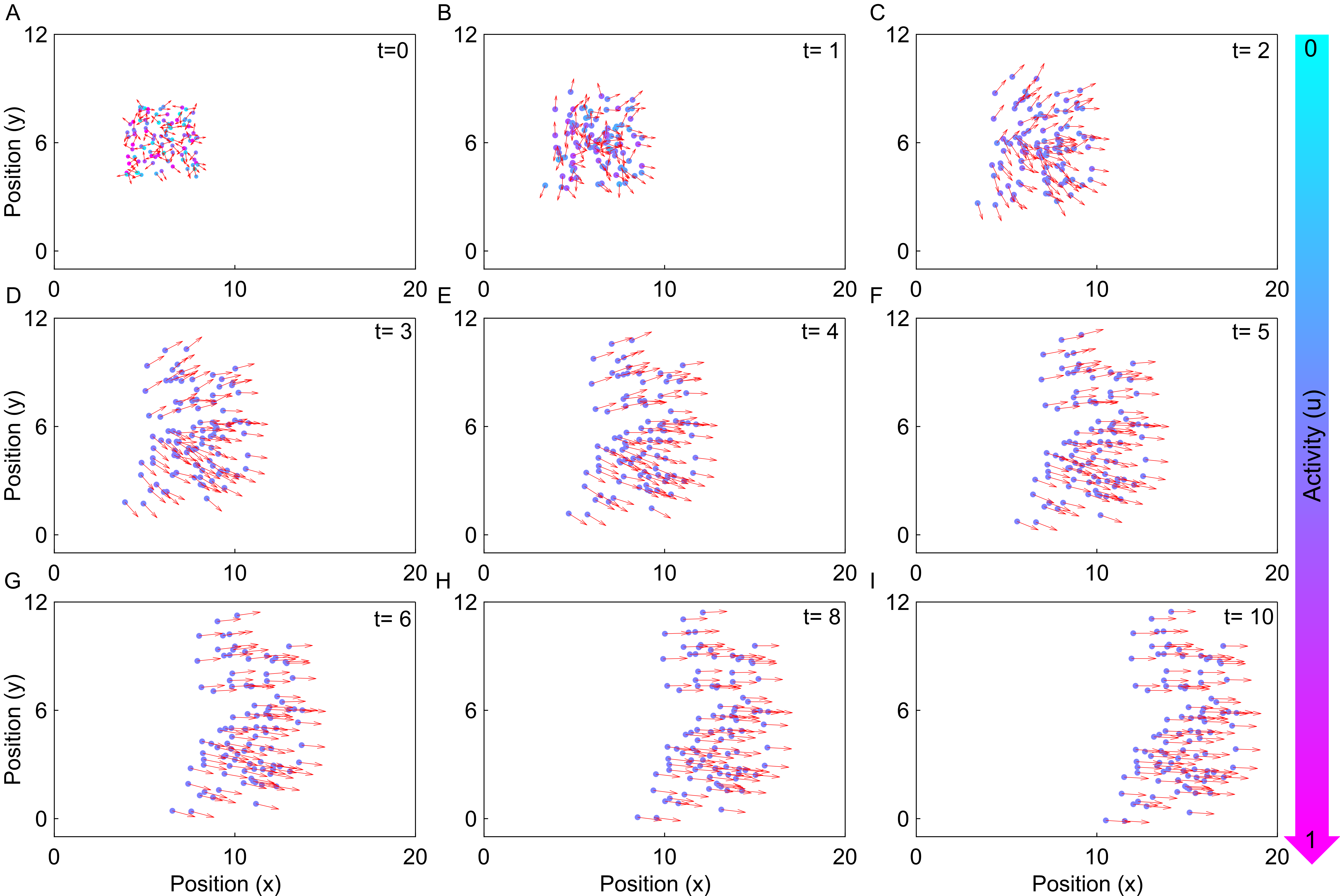} 
   \centering   \vskip-0.3cm
\caption{ \textbf{Spatiotemporal evolution towards consensus in a behavioural swarm.}
Snapshots illustrating the emergent dynamics of $N=100$ agents governed by the consensus model \eqref{test3}, with agents initially distributed in the domain $[4,8] \times [4,8]$. Each agent's state is represented by its position $\bm{x}_i$ (dot), heading angle $\theta_i$ (arrow), and internal activity $u_i \in [0,1]$ (colour scale). The sequence of panels (\textbf{A}-\textbf{I}) shows the progression from random initial conditions (\textbf{A}, $t=0$), characterised by dispersed headings and heterogeneous activities, to a state of global consensus. The system rapidly self-organises into a cohesive, polarised flock, where headings align with the target direction $\theta^\nu=0$ and activities synchronise (\textbf{H}, \textbf{I}). Parameters for this simulation are: $\beta_i = \gamma_i = 1$ and an all-to-all interaction topology.}
\label{fig:t2}
\end{figure}

\begin{figure}[htb!]
  \centering
\includegraphics[width=\textwidth]{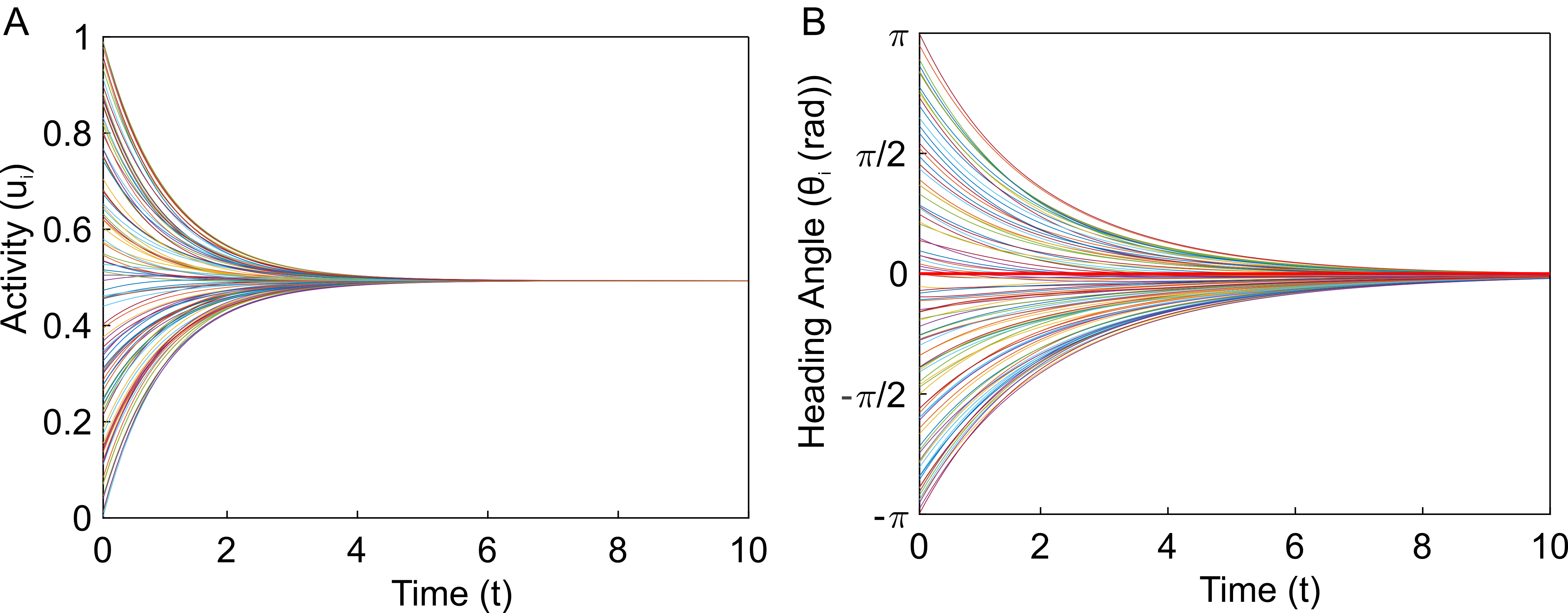} 
\centering   \vskip-0.3cm
\caption{\textbf{Temporal dynamics of state convergence in a behavioural swarm.} Time evolution of the state variables for the $N=100$ agents governed by the consensus model \eqref{test3}. \textbf{(A)} The internal activity $u_i$ for each agent, starting from random values in $[0,1]$, rapidly synchronises to a common equilibrium value. \textbf{(B)} Simultaneously, the heading angle $\theta_i$, initially distributed in $[-\pi, \pi]$, converges unconditionally to the target direction $\theta^\nu=0$. Together, the panels illustrate the robust emergence of collective order.}\label{fig:t3}
\end{figure}

The numerical results of this simulation are presented in Figures~\ref{fig:t2} and \ref{fig:t3}, which collectively illustrate the model's strong tendency towards global consensus. The spatio-temporal evolution depicted in Fig.~\ref{fig:t2} demonstrates this with visual clarity, showing a rapid and robust transition from a disordered initial configuration to a cohesive, polarised flock translating with a common velocity vector $\bm{v}^*= (\cos\theta^\nu,\sin\theta^\nu)$. The initial, high-entropy state, characterised by randomly distributed headings and heterogeneous activities (\textbf{A}), gives way to a globally ordered structure where all agents share a common heading and a synchronised internal state (\textbf{H}, \textbf{I}). These qualitative observations are confirmed with quantitative precision by the temporal dynamics of the state variables presented in Fig.~\ref{fig:t3}. Panel~(\textbf{A}) thereof shows the swift convergence of the individual activities, $u_i(t)$, to a single global consensus value $u^* = \frac{1}{N}\sum_{i=1}^N u_i(0)$, which is the conserved quantity for the linear consensus protocol governing the activity. Simultaneously, panel~(\textbf{B}) illustrates the unconditional alignment of the heading angles, such that as $t \to \infty$, $\theta_i(t) \to \theta^\nu = 0$ for all $i=1,\dots,N$. The uniform, exponential decay towards the equilibrium is characteristic of a stable, linear consensus process, whose rate is governed by the spectral properties of the interaction graph.

However, the assumption of global, unconditional interaction is often a strong idealisation. It is well-established, particularly in socio-economic contexts~\cite{[BL12]}, that interactions can be strongly selective or homophilous. To model this, we introduce a bounded-confidence mechanism, where agents only interact if their internal states are sufficiently close. This is achieved by modifying the communication kernel $\psi$ in~\eqref{test3} with a cut-off function, $\chi(|u_i - u_j| \leq \varepsilon)$, which equals unity if the condition is met and zero otherwise. This yields a new system:

\begin{equation}\label{test4}
\begin{cases}
\frac{du_i}{dt}=  \beta_{i} \sum_{j \in \Omega_i } \chi(|u_i - u_j| \leq \varepsilon) \psi(\|\bm{x}_j -\bm{x}_i\|)
(u_j - u_i ),  \\[0.4cm]
\frac{d\bm{x}_i}{dt}=(\cos{\theta_i},\sin{\theta_i}),\\[0.4cm]
\frac{d\theta_i}{dt} = \gamma_i (\bar\theta_i -\theta_i), \quad 
\bar\theta_i = u_i \theta^\nu + (1-u_i)\sum_{j\in\Omega_i}{\phi(\|\bm{x}_j-\bm{x}_i\|)} \theta_j .
\end{cases}
\end{equation}

\begin{figure}[htb!]
\centering
\includegraphics[width=\textwidth]{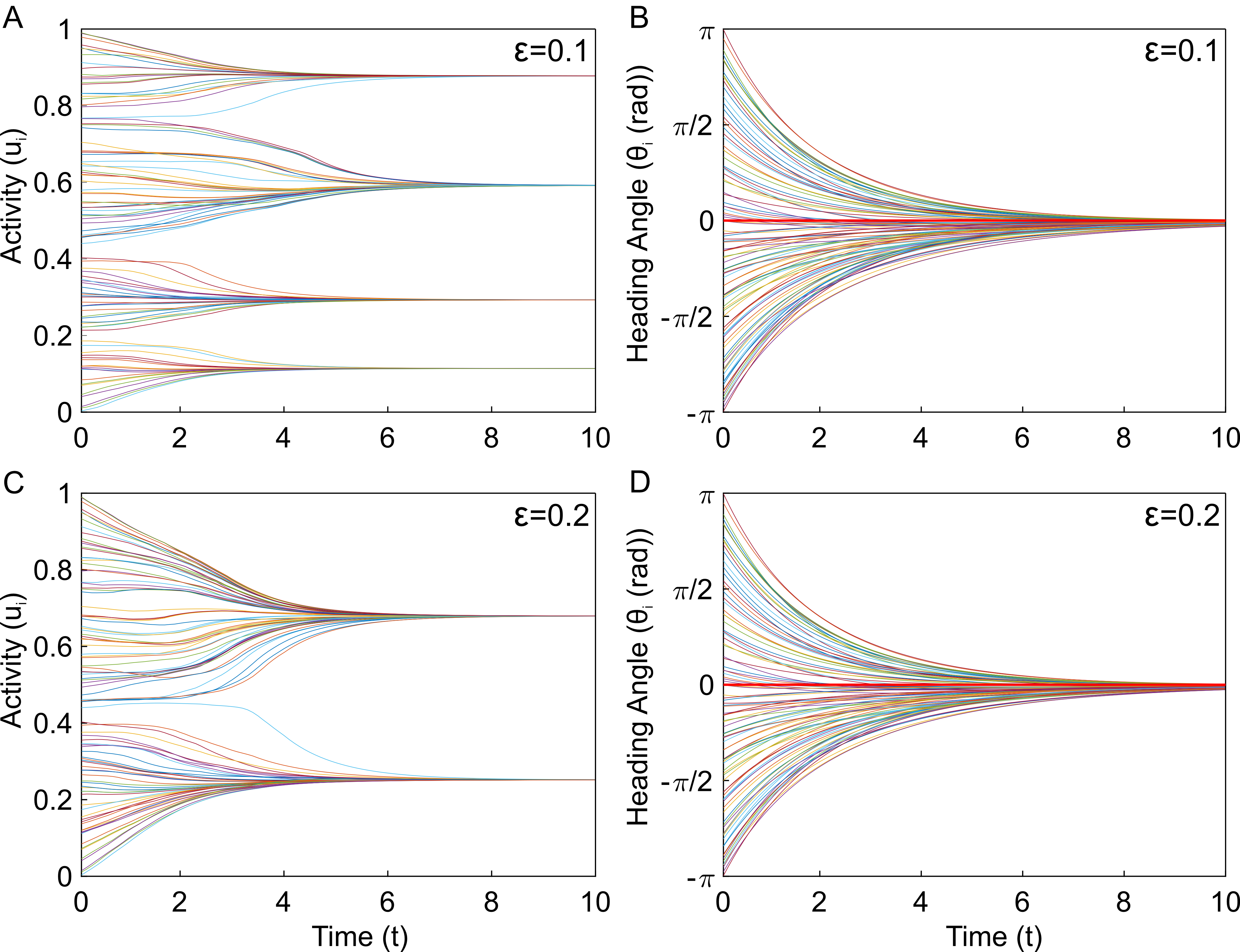} 
\caption{ \textbf{Influence of homophilous interactions on state dynamics and consensus.} Comparative analysis of the multi-consensus model \eqref{test4} under different interaction thresholds, $\varepsilon$. A first simulation with threshold $\varepsilon=0.1$ shows that agent activities $u_i$ fragment into distinct, stable clusters (\textbf{A}), while the corresponding heading angles $\theta_i$ converge to a single global consensus (\textbf{B}). A second simulation using a larger threshold, $\varepsilon=0.2$, presents a similar outcome: the activity again forms clusters (\textbf{C}), and the heading angles converge to the same unified state (\textbf{D}).}
\label{fig:clustering_dynamics}
\end{figure}

The consequences of this state-dependent interaction are explored numerically in Fig.~\ref{fig:clustering_dynamics}, which presents a direct comparison of the system's behaviour for two distinct values of the confidence threshold, $\varepsilon$. The mathematical origin of the observed phenomena lies in the asymmetric modification of the interaction protocols. In the dynamics of the activity, $\frac{du_i}{dt}$, the interaction is now gated by the homophily condition, $\chi(|u_i - u_j| \le \varepsilon)$. This term dynamically severs the links between agents with disparate internal states, leading to a fragmentation of the interaction graph into disconnected components. As seen in panel~(\textbf{A}) for a narrow threshold of $\varepsilon=0.1$, this mathematical fragmentation has a direct physical consequence: the system self-organises into $N_c \approx 5$ distinct, stable opinion blocs, each converging to a local consensus value $u_k^*$.

In striking contrast, the dynamics of the heading angle, $\frac{d\theta_i}{dt}$, remain structurally robust to this fragmentation. The crucial term governing collective alignment, $\theta^S = \sum_{j\in\Omega_i} \phi(\|\bm{x}_j-\bm{x}_i\|)\,\theta_j$, is \textit{not} subject to the homophilous  cut-off. Physically, this implies that while agents are selective about whose `opinion' ($u_j$) they are influenced by, they continue to coordinate their physical `action' ($\theta_j$) with all neighbours in their spatial vicinity. The result, confirmed in panel~(\textbf{B}), is an unconditional convergence to a global kinematic consensus, $\theta_i \to \theta^\nu = 0$.

Increasing the threshold to $\varepsilon=0.2$ (lower panels) corresponds to a more `open-minded' interaction rule. This leads to a coarser-grained fragmentation in the activity space, with fewer but larger opinion blocs ($N_c \approx 3$ in panel~(\textbf{C})), yet the principle of global kinematic consensus remains intact (panel~(\textbf{D})). This decoupling of consensus regimes is perhaps the most profound feature of this theoretical framework. It provides a rigorous mathematical mechanism for describing systems where agents can maintain heterogeneous beliefs or internal states while still achieving remarkable coherence in their collective actions.

\section{Critical analysis and research perspectives}\label{sec:5}

The key motivation behind our essay is the quest for a mathematics that can describe the dynamics of living systems. These systems exhibit behaviours that cannot be captured by classical particle theory. This aspect has already been highlighted in the article on learning intelligence in real swarms, in which each individual selects a fixed number of others to organise their dynamics, see references~\cite{[Ballerini],[CavagnaA]} and~\cite{[CavagnaB]}. The first mathematical formalisation was proposed in~\cite{[BS12]}, which was afterwards technically revisited by various authors. These studies, along with the kinetic theory of active particles reviewed in~\cite{[BBL25]}, have contributed to the development of the theory of behavioural swarms. This theory takes into account the heterogeneity of the activity variable, treating it not as a parameter but as a variable that interacts with all mechanical variables.
\vskip2mm

The specific features of the theory can be summarised as follows: 
\begin{itemize}[itemsep=2mm]
\item Firstly, it provides a formal and rigorous framework explicitly designed for modelling and systematically analysing collective dynamics in active particle systems. This foundational structure is adaptable enough to be applied across various domains, from biological aggregations, such as animal flocks, to engineered systems, including swarms of autonomous robots or software agents.

\item Secondly, a key feature is the theory's inherent ability to manage heterogeneity. It moves beyond the simplifying assumption of identical agents by allowing particles to have diverse physical and behavioural properties and offers mechanisms for these particles to organise into distinct functional subsystems, reflecting the structured complexity of real-world populations.

\item Thirdly, the entire evolution of the system is governed by a sophisticated network of interactions. These are not merely physical forces; they serve as conduits through which an agent's internal behavioural state directly and reciprocally influences its mechanical dynamics. This coupling propels the theory, enabling the modelling of collective behaviours through individual `{\it learning}' and `{\it decision making}', guided by brain-like reasoning and physical constraints. This perspective is supported by the idea that learning through interaction has a significant statistical edge over individual study, a concept notably advanced in the work of Yann LeCun~\cite{[CUN24]} and reinforced by broader research in deep learning~\cite{[BLH21]}.

\item Lastly, the nature of these interactions is fundamentally non-local, extending beyond immediate neighbours to a relevant, yet often limited, subset of the swarm. This recognises that influence in living systems is seldom confined to physical contact, adding realism to the theory. Moreover, the outcomes of these interactions are not predetermined. Instead, they can be described by sophisticated rules, potentially derived from game--theoretic principles, such as the concept of evolutionarily stable strategies~\cite{[MaynardSmith82]}, to capture the complex, strategic behaviours that underpin adaptive collective action.

\end{itemize}

As a first step, a search for research perspectives should be conducted, alongside a comparison with parallel methods that have pursued analogous objectives. This comparison can highlight the strengths of the approach under consideration and identify any limitations that should be addressed for the further development of the theory. The term \textit{parallel methods} refers to mathematical tools designed to describe the collective dynamics of living systems that are governed by differential systems. Specifically, we consider a generalization of the classical kinetic theory, the Kinetic Theory of Active Particles (KTAP) \cite{[BBGO17]}, and methods based on the Fokker--Planck--Boltzmann (FPB) equation~\cite{[PT13]}.

The mathematical theory of behavioural swarms (BST) uses parallel methods to review the merits and drawbacks of BST, focusing mainly on its ability to describe the specific features of living systems, and its flexibility to interact with other differential systems involved in modelling specific systems.

A critical evaluation of the BST highlights several significant strengths that establish it as an important advancement in modelling complex living systems. Chief among these is its impressive ability to capture several essential features of life, a trait it shares with the KTAP, yet it approaches through a different methodological perspective. The theory's strength lies in its treatment of individual strategies and behavioural traits not as fixed, unchanging parameters, but as dynamic variables. This enables the explicit modelling of heterogeneous populations, where behaviours can adapt, evolve, and spread via interactions --- a major shift from models that assume uniformity. Equally important are the sophisticated and realistic interaction mechanisms that BST introduces. It rejects the idea of modelling collective dynamics as a simple linear sum of pairwise, reversible encounters. Instead, it proposes that interactions are inherently non-local, irreversible, and involve multiple particles. This shift recognises that an individual's decision is rarely made in isolation; it is a complex, nonlinear response to the social environment created by a network of interacting agents, thus reflecting a more complete and psychologically plausible reality. Finally, the theory stands out for its notable modelling flexibility, which stems directly from its foundation in systems of ordinary differential equations (ODEs). This mathematically straightforward and well-understood structure offers a manageable and transparent way to connect the swarm's dynamics with various other differential systems. This feature is not a minor convenience; it is essential for addressing the formidable challenge of multiscale modelling~\cite{[Kevrekidis09]}, where phenomena at the cellular, organismic, and population levels must be dynamically and coherently integrated.

Conversely, an honest and forward-looking assessment of the theory must also address its current limitations, which, in turn, outline the most urgent and promising directions for future research. The most immediate and practical challenge is computational scalability. The agent-based, ODE-centric nature of BST, while providing high accuracy for small to medium-sized ensembles, demands substantial computational power as the number of interacting particles grows. This inherent difficulty often leads to a shift towards statistical frameworks, such as KTAP, when dealing with large populations. A second, more significant methodological issue involves the formulation of the interaction functions themselves. Currently, these vital rules are often based on empirical or ad-hoc assumptions, designed to replicate observed phenomena in specific systems. For BST to evolve from a descriptive tool into a genuinely predictive theory, a consistent and general approach for deriving these interaction kernels from first principles---or at least from a minimal set of axioms---must be developed. Most fundamentally, the current structure of BST is inherently number-conservative. The differential system cannot account for proliferative or destructive events, such as interaction-driven birth, death, or state-transition dynamics, which are central to most biological processes, from immune responses to ecosystem development. This limits the theory's applicability in biology, an area where frameworks like KTAP have already made notable progress by including such non-conservative dynamics, for instance in modelling tumour-- immune competition system~\cite{[BD05],[BD06]}. Overcoming this limitation is therefore not merely an extension but also a crucial evolution for the theory to reach its full potential.

All the above considerations suggest that BST differs technically from kinetic theory methods such as KTAP and FPB. These methods require a large number of interacting entities to justify the assumption of continuity of the probability distribution. On the other hand, BST encounters difficulties when the number of particles becomes too large. Nevertheless, the applications reviewed in Section~3 demonstrate that interesting models have already been studied. This suggests exploring further applications for systems that maintain the number of active particles.

 Conversely, applications in various fields necessitate suitable extensions to the existing theory of behavioural swarms (BST). The most challenging aspect is developing mathematical structures that incorporate birth and death dynamics related to interactions. At present, the way in which interactions in BST are modelled is based on empirical rules related to each specific system under consideration. However, a comprehensive theory should adopt a more general approach, inspired by KTAP and guided by the recent paradigm proposed in~\cite{[BBL25]}, where the dynamics are generated by the following sequence:
 \begin{center}
\textit{Physics and Brain Reasoning $\to$  Collective Learning $\to$  Decision Making.}
\end{center}
In detail, the authors of~\cite{[BBL25]} demonstrate how learning and decision-making are guided by a brain-reasoning model that considers the physics of each system. This general concept is applied specifically to modelling human crowds~\cite{[BDL24]}, in which a simple phenomenological model of how the brain works is proposed in the scientific machine learning framework. We bear in mind that these considerations specifically refer to a general differential system that describes the dynamics of each object in the modelling approach. Therefore, it would be interesting to develop the aforementioned reasoning in the structure of the BST.

Beyond the data-driven discovery of interaction forms, the very nature of these interactions opens further research frontiers. For instance, the mathematical description should go beyond consensus dynamics as observed in the book by Bonacich and Lu, see~\cite{[BL12]}. Indeed, recent literature has proposed interesting papers on the interaction between social dynamics, politics, and economics, with examples in~\cite{[DKLM17],[DL14],[DL15]}. We also agree with the idea that the dynamics of interactions, in various cases, involve a combination of different strategies. One example is Parrondo's paradox; see~\cite{[WC24]} and the comment~\cite{[Outada25]}.

The approach developed in~\cite{[BE24]} to derive differential systems that describe the interaction rules that evolve in time within the so-called \textit{Artificial World} in Herbert A.~Simon’s philosophy~\cite{[Simon1965],[Simon2019]} can be studied within the conceptual framework of the mathematical theory of behavioural swarms. In our mind this is the most attractive and challenging perspective.


\subsection*{Data-Driven Perspectives for Behavioural Swarms Theory}

In this way, the synergy between BST and emerging scientific machine learning methods provides a clear path forward. This integration can be pursued through two complementary approaches, which we formalise as the \textit{direct} and \textit{inverse} problems within our theoretical framework.

\subsubsection*{Physics-Informed Inference of Swarm Trajectories}

The core issue involves solving the BST system of ordinary differential equations, often under conditions of sparse or incomplete data that make traditional numerical schemes ill-posed or inaccurate. We suggest addressing this using Physics-Informed Neural Networks (PINNs)~\cite{[Raissi2019jcp]}, a method that transforms the problem of solving a differential system into a variational optimisation framework. The main idea is to approximate the unknown trajectory of each particle's state, $\bm{S}_i(t) = (u_i, z_i, \bm{x}_i, \bm{v}_i)$, using a separate neural network. This network, called $\bm{S}_i^{\text{NN}}$, is a function that takes the time, $t$, as input and provides an approximation of the state vector, $\bm{S}_i$, at that specific time. The network's behaviour is governed by a set of trainable parameters (weights and biases), collectively represented by the vector $\theta_i$. Therefore, the full notation for the network's output is $\bm{S}_i^{\text{NN}}(t; \theta_i)$, which signifies a trial solution from a highly flexible parametric function space.

The core of the method involves restricting this function space to a manifold of physically acceptable solutions. This is accomplished by defining a residual operator, $\mathcal{R}_i$, which functions as a measure of how much the trial solution breaches the BST governing equations:
\[
\mathcal{R}_i(t; \theta) := \frac{d\bm{S}_i^{\text{NN}}(t; \theta)}{dt} - \bm{F}_i(\{\bm{S}_j^{\text{NN}}(t; \theta)\}_{j \in \Omega_i}).
\]
Having established the parametric representation for a single particle, we now consider the entire swarm. The collective set of all trainable parameters for the $N$ particles, denoted $\theta = \{\theta_i\}_{i=1}^N$, is then determined by minimising a composite loss functional, $\mathcal{L}(\theta)$, which combines a data-fidelity term with a physics-based regularisation term:
\begin{equation}
\mathcal{L}(\theta) = \underbrace{\sum_{k \in \text{obs}} \|\bm{S}^{\text{NN}}(t_k; \theta) - \bm{S}_k^{\text{obs}}\|_2^2}_{\text{Data Fidelity Term}} + \underbrace{\lambda\sum_{j \in \text{colloc}} \|\mathcal{R}_j(t_j; \theta)\|_2^2}_{\text{Physical Consistency Term}},
\end{equation}
where $\|\cdot\|_2^2$ denotes the squared $\ell_2$-norm. The total loss is a weighted sum of two components. The first term, the \textit{Data Fidelity Loss}, is calculated over a set of observation points in time, $\{t_k\}_{k \in \text{obs}}$, where the true state of the system, $\bm{S}_k^{\text{obs}}$, is known (e.g., from experimental measurements). The second term, the \textit{Physical Consistency Loss}, is evaluated over a larger set of collocation points, $\{t_j\}_{j \in \text{colloc}}$, sampled from the spatio-temporal domain to ensure the BST equations' validity. The hyperparameter $\lambda > 0$ is a user-defined coefficient that balances the importance of data fitting and physical model compliance. The automatic differentiation features of modern machine learning libraries are essential, as they enable precise calculation of the derivative $\frac{d\bm{S}_i^{\text{NN}}}{dt}$ and the gradients of $\mathcal{L}$ with respect to $\theta$. This approach effectively transforms the ODE system into a high-dimensional optimisation problem, yielding solutions that are both consistent with sparse data and the underlying physical theory.

\subsubsection*{The Inverse Problem: Uncovering BST Interaction Laws}

A more fundamental challenge lies in the formulation of the interaction function $\bm{F}_i$ itself. The inverse problem seeks to determine the mathematical form of $\bm{F}_i$ from observational data. We suggest utilising the Sparse Identification of Nonlinear Dynamics (SINDy) framework~\cite{[Brunton2016pnas]}, which presumes that the governing dynamics are sparse within a high-dimensional space of candidate functions.

Given time-series data of a swarm's state, such as the high-resolution trajectories of starling murmurations~\cite{[Ballerini]}, and its numerically approximated derivatives, $\{\bm{S}_i(t_k), \dot{\bm{S}}_i(t_k)\}$, we construct a candidate function library, $\bm{\Theta}(\{\bm{S}_j\})$. This is a matrix where each column represents a potential basis function for the dynamics, such as multivariate polynomials or trigonometric functions of the states of the interacting particles. The problem is then to find a coefficient matrix $\bm{\Xi}_i$ that satisfies the linear system:
\[
\dot{\bm{S}}_i \approx \bm{\Theta}(\{\bm{S}_j\}_{j \in \Omega_i}) \bm{\Xi}_i.
\]
Without further constraints, this overdetermined system would produce a dense matrix $\bm{\Xi}_i$, leading to an overly complex and hard-to-interpret model. The main idea behind SINDy is to find the sparsest possible solution for $\bm{\Xi}_i$, usually by solving an  $\ell_0$ or $\ell_1$-regularised regression problem. For example, using an $\ell_1$ (LASSO) penalty, the objective becomes:
\begin{equation}
\bm{\Xi}_i^* = \arg\min_{\bm{\Xi}_i} \left( \underbrace{\|\dot{\bm{S}}_i - \bm{\Theta}(\{\bm{S}_j\}) \bm{\Xi}_i\|_2^2}_{\text{Least-Squares Data Fidelity}} + \underbrace{\gamma \|\bm{\Xi}_i\|_1}_{\text{Sparsity-Promoting Regularisation}} \right),
\end{equation}
Here, $\bm{\Xi}_i^*$ denotes the optimal sparse coefficient matrix.  Its non-zero entries indicate the few active terms from the candidate library that best describe the system's dynamics, thereby revealing the parsimonious structure of the underlying differential equation. The regularisation parameter $\gamma$ governs the balance between data accuracy and model sparsity. By solving this optimisation, SINDy systematically prunes the candidate library, unveiling the smallest set of terms needed to represent the observed dynamics. This facilitates the ab-initio discovery of the analytical forms of the interaction kernels ($\eta_{ij}, \chi_{ij}, \psi_{ij}$), offering a well-founded approach for constructing and validating parsimonious models within the BST framework. This enables the initial discovery of the analytical forms of the interaction kernels ($\eta_{ij}, \chi_{ij}, \psi_{ij}$), providing a solid basis for constructing and validating concise models within the BST framework.

In this comprehensive review, we have consolidated the mathematical foundations of BST and established its capacity as a rigorous framework for modelling adaptive living systems. We have systematically progressed from its conceptual origins and core formalism---centred on the dynamic \textit{activity} variable---to a critical evaluation of its applications and current methodological boundaries. This analysis has not been an end in itself, but rather the necessary groundwork to justify and define the next evolutionary stage of the theory. By identifying the empirical formulation of interaction laws as the key limitation, we have established the rationale for a robust synthesis with data-driven methodologies. This manuscript should therefore be viewed not as a historical summary, but as a foundational blueprint. This work solidifies the past to provide a clear, actionable, and rigorous path forward toward a truly predictive science of collective behaviour.


\section*{Conflict of Interest}

The authors declare that they have no conflict of interest.

\section*{Acknowledgments}

Nisrine Outada gratefully acknowledges the support of the University of Granada through the Modeling Nature (MNAT) Research Unit (project QUAL21-011). Jie Liao acknowledges partial support from the National Natural Science Foundation of China (Grant No. 12471211). Rene Fabregas acknowledges partial support from the María Zambrano-Senior grant funded by the Spanish Ministerio de Universidades and Next-Generation EU; Grant C-EXP-265-UGR23 funded by the Consejería de Universidad, Investigación \& Innovación and the ERDF/EU Andalusia Program; Grant PID2022-137228OB-I00 funded by the Spanish Ministerio de Ciencia, Innovación y Universidades (MICIU/AEI/10.13039/501100011033) and the ERDF/EU; and the Modeling Nature (MNAT) Research Unit (project QUAL21-011).




\begin{thebibliography}{100}

\bibitem{[ALBI19]}
\newblock G.~Albi,  N.~Bellomo, L.~Fermo, S.-Y.~Ha, J.~Kim, L.~Pareschi, D.~Poyato, and J.~Soler,
\newblock Traffic, crowds, and swarms: from kinetic theory and multiscale methods to applications and research perspectives,
\newblock \textit{Mathematical Models and Methods in  Applied Sciences}, \textbf{29}, 1901--2005, (2019).

\bibitem{[BCKY19]}
\newblock H.-O.~Bae, S.-Y.Cho, J.~Kim, and S.-B. Yun,
\newblock A kinetic description for the herding behavior in financial market,
\newblock \textit{Journal of Statistical Physics},  \textbf{176(2)}, 398--424, (2019).

\bibitem{[BCLY19]}
\newblock  H.-O.~Bae, S.-Y.~Cho, S.-K.~Lee, and S.-B.~Yun,
\newblock A particle model for herding phenomena induced by dynamic market signals,
\newblock \textit{Journal of Statistical Physics},  \textbf{177(2)}, 365--398, (2019).

\bibitem{[Ballerini]}
\newblock M.~Ballerini, N.~Cabibbo, R.~Candelier, A.~Cavagna, E.~Cisbani, I.~Giardina, V.~Lecomte, A. Orlandi, G. Parisi, A. Procaccini, M. Viale, and V. Zdravkovic,
\newblock Interaction ruling animal collective behavior depends on topological rather than metric distance: evidence from a field study,
\newblock \textit{Proceedings of the National Academy of Sciences},  \textbf{105(4)}, 1232--1237, (2008).

\bibitem{[Bartucci25]}
\newblock C.~Bartucci, J.-M.~Lasry, and P.-L.~Lions,
\newblock A singular infinite dimensional Hamilton-Jacobi-Bellman equation arising from a storage problem,
\newblock  \textit{Mathematical  Models and Methods in Applied Sciences},  \textbf{35(3)}, 703--731, (2025).

\bibitem{[BBGO17]}
\newblock  N.~Bellomo, A.~Bellouquid, L.~Gibelli, and N.~Outada,
\newblock \textbf{  A Quest Towards a Mathematical Theory of Living Systems},
\newblock Modeling and Simulation in Science, Engineering and Technology, Birkh\"auser/Springer, (2017).

\bibitem{[BBD21]}
\newblock N.~Bellomo, D.~Burini, G.~Dosi, L.~Gibelli, D.~A.~Knopoff, N.~Outada, P.~Terna, and M. E.~Virgillito,
\newblock What is life? A perspective of the mathematical kinetic theory of active particles,
\newblock \textit{Mathematical Models and Methods in Applied Sciences}, \textbf{31}, 1821--1866, (2021).

\bibitem{[BBL25]}
\newblock N.~Bellomo, D.~Burini, and J.~Liao,
\newblock New Trends in Kinetic Theory Towards the Complexity of Living Systems,
\newblock \textit{arXiv preprint arXiv:2506.08752v1}, (2025).

\bibitem{[BBO22]}
\newblock  N.~Bellomo, D.~Burini, and N.~Outada,
\newblock  Multiscale models of Covid-19 with mutations and variants,
\newblock \textit{Networks and Heterogeneous Media},  \textbf{17(3)},  293--310, (2022).

\bibitem{[BBST24]}
\newblock N.~Bellomo, D.~Burini, V.~Secchini, and P.~Terna,
\newblock \textbf{Active Particles Methods in Economics},
\newblock Cambridge Elements, Cambridge University Press, (2024).

\bibitem{[BDL24]}
\newblock N.~Bellomo, M.~Dolfin, and J.~Liao,
\newblock  Life and self-organization on the way to artificial intelligence for collective dynamics,
\newblock \textit{Physics of Life Reviews},  \textbf{51}, 1--8,  (2024).

\bibitem{[BE24]}
\newblock N.~Bellomo and M.~Egidi,
\newblock From Herbert A.~Simon's legacy to the evolutionary artificial world with heterogeneous collective behaviors,
\newblock \textit{Mathematical  Models and  Methods in Applied Sciences},   \textbf{34}, 145--180,  (2024).

\bibitem{[BHKL25]}  
 \newblock N. Bellomo, S. Ha, M. Kim, and J. Liao, 
  \newblock
   \newblock Consensus dynamics of behavioral swarm models with random batch interactions and external noises, 
    \newblock {\it preprint}, (2025).

\bibitem{[BHLY24]}
 \newblock N.~Bellomo, S.-Y.~Ha, J.~Liao, and W.~Yoon,
 \newblock  Behavioral swarms: a mathematical theory
 towards swarm intelligence,
  \newblock   \textit{Mathematical Models and Methods in Applied Sciences}, \textbf{ 34}, 2305--2349,  (2024).

\bibitem{[BHO20]}
\newblock N.~Bellomo, S.-Y.~Ha, and N.~Outada,
\newblock Towards a mathematical theory  of behavioral swarms,
\newblock \textit{ESAIM: Control Theory and Variational Calculus}, \textbf{26} 125,  (2020).

\bibitem{[BHOY22]}
\newblock N.~Bellomo, S.-Y.~Ha, N.~Outada, and J. Yoon,
\newblock On the mathematical theory of behavioral swarms emerging collective dynamics,
\newblock \textit{Mathematical Models and Methods in Applied Sciences}, \textbf{32(14)} 2927--2959,  (2022).

\bibitem{[BS12]}
\newblock N.~Bellomo and J.~Soler,
\newblock On the mathematical theory of the dynamics of swarms viewed as a complex system,
\newblock \textit{ Mathematical  Models and Methods in Applied Sciences},  \textbf{22} paper n.~1140006, (2012).

\bibitem{[BD05]}
\newblock A.~Bellouquid and M.~Delitala,
\newblock Mathematical methods and tools of kinetic theory towards modelling complex biological systems,
\newblock \textit{Mathematical  Models and Methods in Applied Sciences}, \textbf{15(11)}, 1639--1666, (2005).

\bibitem{[BD06]}
\newblock A.~Bellouquid and M.~Delitala,
\newblock \textbf{Modelling Complex Biological Systems - A Kinetic Theory Approach},
\newblock Birkh\"auser-Springer, (2006).



\bibitem{[BLH21]}
\newblock Y.~Bengio, Y.~LeCun, and G.~Hinton,
\newblock Deep Learning for AI,
\newblock \textit{Communications of the ACM},  \textbf{64(7)}, 58--65, (2021).


\bibitem{[BW1989]}
\newblock G.~Beni and J.~Wang,
\newblock Swarm intelligence in cellular robotic systems,
\newblock \textit{Proceed. NATO Advanced Workshop on Robots and Biological Systems},  Heidelberg: Springer, 703--712, (1993).

\bibitem{[Ben-Jacob00]}
\newblock E.~Ben-Jacob, I.~Cohen, and H.~Levine,
\newblock Cooperative self-organization of microorganisms,
\newblock \textit{Advances in Physics}, \textbf{49(4)}, 395--554, (2000).

\bibitem{[B3EPT]}
\newblock G.~Bertaglia, A.~Bondesan, D.~Burini, E.~Eftimie, L.~Pareschi, and G.~Toscani,
\newblock  New trends on the systems approach to modeling SARS-CoV-2 pandemics in a globally connected planet,
\newblock \textit{Mathematical Models and Methods in Applied Sciences}, \textbf{34(11)}, 1995--2054, (2024).

\bibitem{[BP21]}
\newblock G.~Bertaglia and L.~Pareschi,
\newblock Hyperbolic compartmental models for epidemic spread on networks with uncertain data: application to the emergence of Covid-19 in Italy,
\newblock \textit{Mathematical  Models and Methods in Applied Sciences},  \textbf{31}, 2495--2531, (2021).

\bibitem{[BMST15]}
\newblock R.~Boero, M.~Morini, M.~Sonnessa, and P.~Terna,
\newblock \textbf{Agent-based Models of the Economy From Theories to Applications},
\newblock Palgrave McMillan, (2015).

\bibitem{[BL12]}
\newblock P.~Bonacich and P.~Lu,
\newblock \textbf{Introduction to Mathematical Sociology},
\newblock Princeton University Press, (2012).

\bibitem{[Brunton2016pnas]}
\newblock S.~L.~Brunton, J.~L.~Proctor, and J.~N.~Kutz,
\newblock Discovering governing equations from data by sparse identification of nonlinear dynamical systems,
\newblock \textit{Proceedings of the National Academy of Sciences}, \textbf{113(15)}, 3932--3937, (2016).

\bibitem{[BK24]}
\newblock D.~Burini and D.~A.~Knopoff,
\newblock Epidemics and society: a multiscale vision from the small world to the globally interconnected world,
\newblock \textit{Mathematical Models and Methods in Applied Sciences}, \textbf{34(8)}, 1564--1594 (2024).


\bibitem{[BDF17]}
\newblock D. Burini, S. De Lillo and G. Fioriti, 
\newblock Influence of drivers ability in a discrete vehicular traﬃc model, 
\newblock \textit{International Journal of Modern Physics}, C \textbf{28(3)}, 1750030, (2017).


\bibitem{[Caponigro17]}
\newblock M.~Caponigro, B.~Piccoli, F.~Rossi, and E.~Trélat,
\newblock Mean-field sparse Jurdjevic-Quinn control,
\newblock \textit{Mathematical Models and Methods in Applied Sciences}, \textbf{27(07)}, 1223-1253, (2017).

\bibitem{[CFL09]}
\newblock C.~Castellano, S.~Fortunato, and V.~Loreto,
\newblock Statistical physics of social dynamics,
\newblock \textit{Reviews of Modern Physics}, \textbf{81}(2), 591--646, (2009).

\bibitem{[Cavagna10]}
\newblock A.~Cavagna, A.~Cimarelli, I.~Giardina, G.~Parisi, R.~Santagati, F.~Stefanini, and M.~Viale,
\newblock Scale-free correlations in starling flocks,
\newblock \textit{Proceedings of the National Academy of Sciences}, \textbf{107(26)}, 11865--11870, (2010).

\bibitem{[CavagnaA]}
\newblock A.~Cavagna, I.~Giardina, A.~Orlandi, G.~Parisi, A.~Procaccini, M.~Viale, and V.~Zdravkovic,
\newblock The STARFLAG handbook on collective animal behaviour: 1. Empirical methods,
\newblock \textit{Animal Behaviour}, \textbf{76}, 217--236, (2008).

\bibitem{[CavagnaB]}
\newblock A.~Cavagna, I.~Giardina, A.~Orlandi, G.~Parisi, A.~Procaccini, M.~Viale, and V.~Zdravkovic,
\newblock The STARFLAG handbook on collective animal behaviour: 2. Three-dimensional analysis,
\newblock \textit{Animal Behaviour}, \textbf{76}, 237--248, (2008).

\bibitem{[CLE19]}
\newblock C.~Cleland,
\newblock  \textbf{The Quest for a Universal Theory of Life, Searching for Life as We Don't Know it},
\newblock  Cambridge University Press, (2019).

\bibitem{[CHL17]}
\newblock Y.-P. Choi, S.-Y. Ha, and Z. Li,
\newblock Emergent Dynamics of the Cucker--Smale Flocking Model and Its Variants,
\newblock In Active Particles, Volume 1: Advances in Theory, Models, and Applications, (2017).

\bibitem{[CDF07]}
\newblock V.~Coscia, M.~Delitala, and P.~Frasca,
\newblock On the mathematical theory of vehicular traffic flow models II: discrete velocity kinetic models,
\newblock   \textit{International Journal Non-linear Mechanics}, \textbf{42}, 411-421, (2007).

\bibitem{[Couzin06]}
\newblock I.~D.~Couzin, J.~Krause, N.~R.~Franks, and S.~A.~Levin,
\newblock Effective leadership and decision-making in animal groups on the move,
\newblock \textit{Nature}, \textbf{433}, 513--516, (2005).

\bibitem{[CK02]}
\newblock I.~D.~Couzin, J.~Krause, R.~James, G.~D.~Ruxton and N.~R.~Franks,
\newblock Collective memory and spatial sorting in animal groups,
\newblock {\it Journal of Theoretical Biology},  {\bf  218}, 1--11,  (2002).

\bibitem{[CH08]}
\newblock F.~Cucker and C.~Huepe,
\newblock Flocking with informed agents,
\newblock \textit{Mathematics in Action}, \textbf{ 1(1)}, 1--25, (2008).

\bibitem{[CS07]}
\newblock F.~Cucker and S.~Smale,
\newblock Emergent behavior in flocks,
\newblock \textit{IEEE Transactions on automatic control}, \textbf{ 52(5)}, 852--862, (2007).

\bibitem{[CS2007]}
\newblock F.~Cucker and S.~Smale,
\newblock On the mathematics of emergence,
\newblock \textit{Japanese Journal of Mathematics}, \textbf{2:197}--227, (2007).

\bibitem{[DKLM17]}
\newblock M.~Dolfin, D.~Knopoff, L.~Leonida, and D.~Patti,
\newblock Escaping the trap of ``blocking'': a kinetic model linking economic development and political competition,
\newblock   \textit{Kinetic and Related Models},   \textbf{10}, 423--443,  (2017).

\bibitem{[DL14]}
\newblock M.~Dolfin and M.~Lachowicz,
\newblock Modeling altruism and selfishness in welfare dynamics: the role of nonlinear interactions,
\newblock \textit{Mathematical  Models and Methods in Applied Sciences}, \textbf{24}, 2361--2381, (2014).

\bibitem{[DL15]}
\newblock M.~Dolfin and M.~Lachowicz,
\newblock Modeling opinion dynamics: how the network enhances consensus,
\newblock  \textit{Networks Heterogeneous Media},  \textbf{10(4)},  421--441, (2015).

\bibitem{[DLO17]}
\newblock M.~Dolfin,  L.~Leonida, and N.~Outada,
\newblock  Modelling human behaviour in economics and social science,
\newblock  \textit{Physics of Life Reviews}, \textbf{22}, 1--21, (2017).

\bibitem{[FPTT17]}
\newblock  G.~Furioli, A.~Pulvirenti, E.~Terraneo, and  G.~Toscani,
\newblock Fokker-Planck equations in the modeling of socio-economic phenomena,
\newblock   \textit{Mathematical  Models and Methods in Applied Sciences}, \textbf{27}, 115--158, (2017).

\bibitem{[FPTT20]}
\newblock G.~Furioli, A.~Pulvirenti, E.~Terraneo, and G.~Toscani,
\newblock Non-Maxwellian kinetic equations modeling the dynamics of wealth distribution,
\newblock  \textit{Mathematical  Models and Methods in Applied Sciences}, \textbf{30}, 685--725, (2020).

\bibitem{[HKR18]}
\newblock S.-Y.~Ha, J.~Kim, and T.~Ruggeri,
\newblock Emergent behaviors of thermodynamic Cucker--Smale particles,
\newblock \textit{SIAM Journal on Mathematical Analysis}, \textbf{50}, 3092--3121, (2018).

\bibitem{[HLL09]}
\newblock S.-Y.~Ha, K.~Lee, and D.~Levy,
\newblock Emergence of time-asymptotic flocking in a stochastic Cucker--Smale system,
\newblock \textit{Communications in Mathematical Sciences}, \textbf{ 7(2)}, 453--469, (2009).

\bibitem{[HK02]}
\newblock R.~Hegselmann and U.~Krause,
\newblock Opinion dynamics and bounded confidence models,
\newblock \textit{Journal of Artificial Societies and Social Simulation}, \textbf{5}(3), (2002).

\bibitem{[HR17]}
\newblock S.-Y.~Ha and T.~Ruggeri,
\newblock Emergent dynamics of a thermodynamically consistent particle model,
\newblock \textit{Archive for Rational Mechanics and Analysis}, \textbf{ 223}, 1397--1425, (2017).

\bibitem{[Hartwell]}
\newblock H.L.~Hartwell, J.J.~Hopfield, S.~Leibler, and A.W.~Murray,
\newblock From molecular to modular cell biology,
\newblock  \textit{Nature}, \textbf{402}  c47--c52, (1999).

\bibitem{[Helbing95]}
\newblock D.~Helbing and P.~Moln\'ar,
\newblock Social force model for pedestrian dynamics,
\newblock \textit{Physical Review E}, \textbf{51(5)}, 4282--4286, (1995).

\bibitem{[Jadbabaie03]}
\newblock A.~Jadbabaie, J.~Lin, and A.~S.~Morse,
\newblock Coordination of groups of mobile autonomous agents using nearest neighbor rules,
\newblock \textit{Proceedings of the 41st IEEE Conference on Decision and Control, 2002.}, \textbf{3}, 2953-2958, (2002).

\bibitem{[Karniadakis21]}
\newblock G.~E.~Karniadakis, I.~G.~Kevrekidis, L.~Lu, P.~Perdikaris, S.~Wang, and L.~Yang,
\newblock Physics-informed machine learning,
\newblock \textit{Nature Reviews Physics}, \textbf{3(6)}, 422--440, (2021).

\bibitem{[Kevrekidis09]}
\newblock I.~G.~Kevrekidis and G.~Samaey,
\newblock Equation-free multiscale computation: algorithms and applications,
\newblock \textit{Annual Review of Physical Chemistry}, \textbf{60}, 321--344, (2009).

\bibitem{[KLY25]}
\newblock D.A.~Knopoff, J.~Liao, Q.~Ma, and X.~Yang,
\newblock Individual-based crowd dynamics with social interaction,
\newblock \textit{Mathematical  Models and Methods in Applied Sciences}, to appear, (2025).

\bibitem{[KST20]}
\newblock D.~Knopoff, V.~Secchini, and  P.~Terna,
\newblock Cherry picking: consumer choices in swarm dynamics, considering price and quality of goods,
\newblock \textit{Symmetry}, \textbf{12(11)}, 1912, (2020).

\bibitem{[KKHP]}
\newblock H.L.~Kwa, J.L.~Kit, N.~Horsevad, J.~Philippot, M.~Savari, and R.~Bouffanais,
\newblock Adaptivity: a path towards general swarm intelligence?
\newblock \textit{Frontiers Robot. AI}, 10:1163185, (2023).
\newblock  doi: 10.3389/frobt.2023.1163185

\bibitem{[LL07]}
\newblock J.-M.~Lasry and P.-L.~Lions,
\newblock Mean field games,
\newblock  \textit{Japanese Journal of Mathematics}, \textbf{2(1)}, 229--260, (2007).



\bibitem{[CUN24]}
\newblock Y.~LeCun,
\newblock Il manque aux machines le sens commun,
\newblock \textit{La Recherche},  \textbf{Avril/Juin}, 20--23, (2024).


\bibitem{[Li14]}
\newblock Z.~Li,
\newblock Effectual leadership in flocks with hierarchy and individual preference,
\newblock \textit{Discrete and Continuous Dynamical Systems}, \textbf{34(9)}, 3683--3702, (2014).

\bibitem{[LH15]}
\newblock Z.~Li and S.-Y.~Ha,
\newblock On the Cucker--Smale flocking with alternating leaders,
\newblock \textit{Quarterly of Applied Mathematics}, \textbf{ 73(4)}, 693--709, (2015).

\bibitem{[LX10]}
\newblock Z.~Li and X.~Xue,
\newblock Cucker--Smale flocking under rooted leadership with fixed and switching topologies,
\newblock \textit{SIAM Journal on Applied Mathematics}, \textbf{70(8)}, 3156--3174, (2010).

\bibitem{[Marchetti13]}
\newblock M.~C.~Marchetti, J.~F.~Joanny, S.~Ramaswamy, T.~B.~Liverpool, J.~Prost, M.~Rao, and R.~A.~Simha,
\newblock Hydrodynamics of soft active matter,
\newblock \textit{Reviews of Modern Physics}, \textbf{85(3)}, 1143--1189, (2013).

\bibitem{[MAY]}
\newblock R.M.~May,
\newblock Uses and abuses of mathematics in biology,
\newblock \textit{Science}, \textbf{303}, 338--342, (2004).

\bibitem{[MaynardSmith82]}
\newblock J.~Maynard Smith,
\newblock \textbf{Evolution and the Theory of Games},
\newblock Cambridge University Press, Cambridge, (1982).

\bibitem{[MMT19]}
\newblock M.~Mazzoli, M.~Morini, and P.~Terna,
\newblock \textbf{Rethinking Macroeconomics with Endogenous Market Structure},
\newblock \textit{Cambridge University Press}, (2019).

\bibitem{[MT11]}
\newblock S.~Motsch and E.~Tadmor,
\newblock A new model for self-organized dynamics and its flocking behavior,
\newblock \textit{ Journal of Statistical Physics}, \textbf{144}, 923--947, (2011).

\bibitem{[MT11B]}
\newblock S.~Motsch and E.~Tadmor,
\newblock Heterophilious dynamics enhances consensus,
\newblock \textit{SIAM Review}, \textbf{56(4)}, 577--621, (2014).

\bibitem{[Nowak]}
\newblock M.A.~Nowak,
\newblock \textbf{Evolutionary Dynamics: Exploring the Equations of Life},
\newblock Harvard University Press, Cambridge (MA), (2006).

\bibitem{[NM01]}
\newblock M.A.~Nowak and R.~May,
\newblock  \textbf{Virus Dynamics: Mathematical Principles of Immunology and Virology},
\newblock Oxford University Press, Oxford, (2001).

\bibitem{[Outada25]}
\newblock N.~Outada,
\newblock Reasonings on multiple strategies in differential systems: comment on ``Parrondo's paradox reveals counterintuitive wins in biology and decision making in society'' by T. Wen \& K.H. Cheong,
\newblock \textit{Physics of Life Reviews}, \textbf{52}, 248--249, (2025).

\bibitem{[PT13]}
\newblock L.~Pareschi and G.~Toscani,
\newblock  \textbf{Interacting Multiagent Systems: Kinetic Equations and Monte Carlo Methods},
\newblock Oxford University Press, Oxford, (2013).

\bibitem{[Parisi23]}
\newblock G.~Parisi,
\newblock Nobel Lecture: multiple equilibria,
\newblock \textit{Reviews of Modern Physics}, \textbf{95(3)}, 030501, (2023).

\bibitem{[Ilia]}
\newblock I.~Prigogine and R.~Herman,
\newblock  \textbf{Kinetic Theory of Vehicular Traffic},
\newblock Elsevier, New York, (1971).

\bibitem{[REED]}
\newblock R.~Reed,
\newblock Why is mathematical biology so hard?,
\newblock \textit{ Notices of the American Mathematical Society}, \textbf{51}, 338--342, (2004).

\bibitem{[RAF19]}
\newblock S.~M.~Reia, A.~C.~Amado, and J.~F.~Fontanari,
\newblock Agent-based models of collective intelligence,
\newblock \textit{Physics of Life Reviews},  \textbf{31}, 320--331, (2019).

\bibitem{[Raissi2019jcp]}
\newblock M.~Raissi, P.~Perdikaris, and G.~E.~Karniadakis,
\newblock Physics-informed neural networks: A deep learning framework for solving forward and inverse problems involving nonlinear partial differential equations,
\newblock \textit{Journal of Computational Physics}, \textbf{378}, 686--707, (2019).

\bibitem{[Schrodinger]}
\newblock  E.~Schr\"odinger,
\newblock \textbf{What is Life? The Physical Aspect of the Living Cell},
\newblock Cambridge University Press, Cambridge, (1944).

\bibitem{[S2008]}
\newblock J.~Shen,
\newblock Cucker--Smale flocking under hierarchical leadership,
\newblock \textit{SIAM Journal on Applied Mathematics}, \textbf{68(3)}:694--719, (2008).

\bibitem{[Simon1965]}
\newblock H.A.~Simon,
\newblock The architecture of complexity,
\newblock \textit{General systems},  \textbf{10},  63--76, (1965).

\bibitem{[Simon2019]}
\newblock H.~A.~Simon,
\newblock \textbf{The Science of the Artificial}, Third Edition,
\newblock MIT Press, Boston, (2019).

\bibitem{[Toner95]}
\newblock J.~Toner and Y.~Tu,
\newblock Long-range order in a two-dimensional dynamical XY model: how birds fly together,
\newblock \textit{Physical Review Letters}, \textbf{75(23)}, 4326--4329, (1995).

\bibitem{[Vicsek95]}
\newblock T.~Vicsek, A.~Czir\'ok, E.~Ben-Jacob, I.~Cohen, and O.~Shochet,
\newblock Novel type of phase transition in a system of self-driven particles,
\newblock \textit{Physical Review Letters}, \textbf{75(6)}, 1226--1229, (1995).

\bibitem{[vonNeumann66]}
\newblock J.~von Neumann,
\newblock \textbf{Theory of Self-Reproducing Automata},
\newblock University of Illinois Press, Urbana, (1966).



\bibitem{[WC24]}
\newblock T.~Wen and K.-H.~Cheong,
\newblock Parrondo's paradox reveals counterintuitive wins in biology and decision making in society,
 \newblock \textit{Physics of Life Reviews},  \textbf{51}:33--59,  (2024).



\bibitem{[Zagour23]}
\newblock M.~Zagour,
\newblock Modeling and numerical simulations of multilane vehicular traffic by active particles methods,
\newblock \textit{Mathematical Models and Methods in Applied Sciences}, \textbf{33(05)}, 1119--1146, (2023).

\bibitem{[Zhang]}
\newblock W.-B. Zhang,
\newblock \textbf{Chaos, Complexity, and Nonlinear Economic Theory},
\newblock World Scientific, (2023).

















\end{thebibliography}
\end{document}